\documentclass[aps,prc,amsfonts,superscriptaddress,twocolumn,showpacs,nofootinbib,showkeys]{revtex4-1}  
\usepackage{graphicx}  
\usepackage{amssymb}   
\usepackage{amsmath}
\usepackage{hyperref}
\usepackage{verbatim}
\usepackage{multirow}
\usepackage[modulo]{lineno}

\begin{document}

\title{High-resolution, accurate MR-TOF-MS for short-lived, exotic nuclei of few events in their ground and low-lying isomeric states}



\author{Samuel Ayet San Andr\'{e}s}
\affiliation{II.~Physikalisches Institut, Justus-Liebig-Universit{\"a}t Gie{\ss}en, 35392 Gie{\ss}en, Germany}
\affiliation{GSI Helmholtzzentrum f{\"u}r Schwerionenforschung GmbH, 64291 Darmstadt, Germany}
\author{Christine Hornung}
\affiliation{II.~Physikalisches Institut, Justus-Liebig-Universit{\"a}t Gie{\ss}en, 35392 Gie{\ss}en, Germany}
\author{Jens Ebert}
\affiliation{II.~Physikalisches Institut, Justus-Liebig-Universit{\"a}t Gie{\ss}en, 35392 Gie{\ss}en, Germany}

\author{Wolfgang R.\ Pla\ss}
\affiliation{II.~Physikalisches Institut, Justus-Liebig-Universit{\"a}t Gie{\ss}en, 35392 Gie{\ss}en, Germany}
\affiliation{GSI Helmholtzzentrum f{\"u}r Schwerionenforschung GmbH, 64291 Darmstadt, Germany}
\author{Timo Dickel}
\email[Corresponding author: ]{t.dickel@gsi.de}
\affiliation{II.~Physikalisches Institut, Justus-Liebig-Universit{\"a}t Gie{\ss}en, 35392 Gie{\ss}en, Germany}
\affiliation{GSI Helmholtzzentrum f{\"u}r Schwerionenforschung GmbH, 64291 Darmstadt, Germany}
\author{Hans Geissel}
\affiliation{II.~Physikalisches Institut, Justus-Liebig-Universit{\"a}t Gie{\ss}en, 35392 Gie{\ss}en, Germany}
\affiliation{GSI Helmholtzzentrum f{\"u}r Schwerionenforschung GmbH, 64291 Darmstadt, Germany}
\author{Christoph~Scheidenberger}
\affiliation{II.~Physikalisches Institut, Justus-Liebig-Universit{\"a}t Gie{\ss}en, 35392 Gie{\ss}en, Germany}
\affiliation{GSI Helmholtzzentrum f{\"u}r Schwerionenforschung GmbH, 64291 Darmstadt, Germany}

\author{Julian Bergmann}
\affiliation{II.~Physikalisches Institut, Justus-Liebig-Universit{\"a}t Gie{\ss}en, 35392 Gie{\ss}en, Germany}
\author{Florian Greiner}
\affiliation{II.~Physikalisches Institut, Justus-Liebig-Universit{\"a}t Gie{\ss}en, 35392 Gie{\ss}en, Germany}
\author{Emma Haettner}
\affiliation{GSI Helmholtzzentrum f{\"u}r Schwerionenforschung GmbH, 64291 Darmstadt, Germany}
\author{Christian Jesch}
\affiliation{II.~Physikalisches Institut, Justus-Liebig-Universit{\"a}t Gie{\ss}en, 35392 Gie{\ss}en, Germany}
\author{Wayne Lippert}
\affiliation{II.~Physikalisches Institut, Justus-Liebig-Universit{\"a}t Gie{\ss}en, 35392 Gie{\ss}en, Germany}
\author{Israel Mardor}
\affiliation{Tel Aviv University, 6997801 Tel Aviv, Israel}
\affiliation{Soreq Nuclear Research Center, 81800 Yavne, Israel}
\author{Ivan Miskun}
\affiliation{II.~Physikalisches Institut, Justus-Liebig-Universit{\"a}t Gie{\ss}en, 35392 Gie{\ss}en, Germany}
\author{Zygmunt Patyk}
\affiliation{National Centre for Nuclear Research, Hoża 69, 00-681 Warszawa, Poland}
\author{Stephane Pietri}
\affiliation{GSI Helmholtzzentrum f{\"u}r Schwerionenforschung GmbH, 64291 Darmstadt, Germany}
\author{Alexander Pihktelev} 
\affiliation{Institute for Energy Problems of Chemical Physics, RAS, 142432 Chernogolovka - Moscow, Russia}
\author{Sivaji Purushothaman}
\affiliation{GSI Helmholtzzentrum f{\"u}r Schwerionenforschung GmbH, 64291 Darmstadt, Germany}
\author{Moritz P.~Reiter} 
\affiliation{II.~Physikalisches Institut, Justus-Liebig-Universit{\"a}t Gie{\ss}en, 35392 Gie{\ss}en, Germany}
\affiliation{TRIUMF, BC V6T 2A3 Vancouver, Canada}
\author{Ann-Kathrin~Rink}
\affiliation{II.~Physikalisches Institut, Justus-Liebig-Universit{\"a}t Gie{\ss}en, 35392 Gie{\ss}en, Germany}
\author{Helmut Weick}
\affiliation{GSI Helmholtzzentrum f{\"u}r Schwerionenforschung GmbH, 64291 Darmstadt, Germany}
\author{Mikhail I.\ Yavor} 
\affiliation{Institute for Analytical Instrumentation, RAS, 190103 St. Petersburg, Russia}

\author{Soumya Bagchi} 
\affiliation{II.~Physikalisches Institut, Justus-Liebig-Universit{\"a}t Gie{\ss}en, 35392 Gie{\ss}en, Germany}
\affiliation{GSI Helmholtzzentrum f{\"u}r Schwerionenforschung GmbH, 64291 Darmstadt, Germany}
\affiliation{Saint Mary's University, NS B3H 3C3 Halifax, Canada}
\author{Volha Charviakova}
\affiliation{National Centre for Nuclear Research, Hoża 69, 00-681 Warszawa, Poland}
\author{Paul Constantin} 
\affiliation{IFIN-HH/ELI-NP, 077126, M\u{a}gurele - Bucharest, Romania}
\author{Marcel Diwisch}
\affiliation{II.~Physikalisches Institut, Justus-Liebig-Universit{\"a}t Gie{\ss}en, 35392 Gie{\ss}en, Germany}
\author{Andrew Finlay} 
\affiliation{TRIUMF, BC V6T 2A3 Vancouver, Canada}
\author{Satbir Kaur}
\affiliation{Saint Mary's University, NS B3H 3C3 Halifax, Canada}
\author{Ronja Kn{\"o}bel}
\affiliation{GSI Helmholtzzentrum f{\"u}r Schwerionenforschung GmbH, 64291 Darmstadt, Germany}
\author{Johannes Lang}
\affiliation{II.~Physikalisches Institut, Justus-Liebig-Universit{\"a}t Gie{\ss}en, 35392 Gie{\ss}en, Germany}
\author{Bo Mei}
\affiliation{IFIN-HH/ELI-NP, 077126, M\u{a}gurele - Bucharest, Romania}
\author{Iain D.\ Moore}
\affiliation{University of Jyv{\"a}skyl{\"a}, 40014 Jyv{\"a}skyl{\"a}, Finland}
\author{Jan-Hendrik Otto}
\affiliation{II.~Physikalisches Institut, Justus-Liebig-Universit{\"a}t Gie{\ss}en, 35392 Gie{\ss}en, Germany}
\author{Ilkka Pohjalainen}
\affiliation{University of Jyv{\"a}skyl{\"a}, 40014 Jyv{\"a}skyl{\"a}, Finland}
\author{Andrej Prochazka}
\affiliation{GSI Helmholtzzentrum f{\"u}r Schwerionenforschung GmbH, 64291 Darmstadt, Germany}
\author{Christophe Rappold}
\affiliation{II.~Physikalisches Institut, Justus-Liebig-Universit{\"a}t Gie{\ss}en, 35392 Gie{\ss}en, Germany}
\affiliation{GSI Helmholtzzentrum f{\"u}r Schwerionenforschung GmbH, 64291 Darmstadt, Germany}
\author{Maya Takechi}
\affiliation{GSI Helmholtzzentrum f{\"u}r Schwerionenforschung GmbH, 64291 Darmstadt, Germany}
\author{Yoshiki K.\ Tanaka}
\affiliation{GSI Helmholtzzentrum f{\"u}r Schwerionenforschung GmbH, 64291 Darmstadt, Germany}
\author{John S.\ Winfield}
\affiliation{GSI Helmholtzzentrum f{\"u}r Schwerionenforschung GmbH, 64291 Darmstadt, Germany}
\author{Xiaodong Xu}
\affiliation{II.~Physikalisches Institut, Justus-Liebig-Universit{\"a}t Gie{\ss}en, 35392 Gie{\ss}en, Germany}

\date{\today}
             
\begin{abstract}

Mass measurements of fission and projectile fragments, produced via $^{238}$U and $^{124}$Xe primary beams, have been performed with the multiple-reflection time-of-flight mass spectrometer (MR-TOF-MS) of the FRS Ion Catcher with a mass resolving powers (FWHM) up to 410,000 and an uncertainty of $6\cdot 10^{-8}$. The nuclides were produced and separated in-flight with the fragment separator FRS at 300 to 1000 MeV/u and thermalized in a cryogenic stopping cell. The data-analysis procedure was developed to determine with highest accuracy the mass values and the corresponding uncertainties for the most challenging conditions: down to a few events in a spectrum and overlapping distributions, characterized only by a broader common peak shape. With this procedure, the resolution of low-lying isomers is increased by a factor of up to three compared to standard data analysis. The ground-state masses of 31 short-lived nuclides of 15 different elements with half-lives down to 17.9~ms and count rates as low as 11 events per nuclide were determined. This is the first direct mass measurement for seven nuclides. The excitation energies and the isomer-to-ground state ratios of six isomeric states with excitation energies down to about 280~keV were measured. For nuclides with known mass values, the average relative deviation from the literature values is $(2.9 \pm 6.2) \cdot 10^{-8}$. The measured two-neutron separation energies and their slopes near and at the N=126 and Z=82 shell closures indicate a strong element-dependent binding energy of the first neutron above the closed proton shell Z=82. The experimental results deviate strongly from the theoretical predictions, especially for N=126 and N=127. 
\end{abstract}


\maketitle


\section{Introduction}
\label{S:1}
Masses are a key property of atomic nuclei. Accurate measurements are needed to understand the evolution of nuclear structure \cite{bohr1998} and stellar nucleosynthesis \cite{Schatz2013}. In particular, nuclear masses indicate the limits of nuclear existence, changes in nuclear deformation and the onset of nuclear collectivity \cite{zhang2016}. Accurate mass values are an important nuclear ingredient to r-process calculations \cite{Mumpower2016}. They significantly affect the description of the equation-of-state of nuclear matter, which can be extended to describe neutron-star matter and crustal composition \cite{wolf2013}.

The knowledge of isomer excitation energies and isomer-to-ground state ratios are of great importance to nuclear structure and reactions. Direct measurements of excitation energies can be complementary to gamma de-excitation measurements. Mass measurements are for long-lived isomers the only applicable method.

Driven by this motivation, multiple-reflection time-of-flight mass spectrometry \cite{Wollnik1990} has been developed to determine nuclear masses of very exotic nuclei at ground and isomeric states, which have half-lives as short as a few milliseconds and which can only be produced with a few events per hour or day \cite{Plass2013b}. It has a unique combination of performance parameters: fast (cycle times of a few milliseconds), accurate (relative mass measurement uncertainty below $10^{-6}$), sensitive (only a few detected ions per nuclide are required for the accurate mass determination) and non-scanning (simultaneous measurement of many different nuclides). Established methods for mass measurements of exotic nuclei, such as isochronous \cite{Knobel2016} or Schottky mass spectrometry \cite{Litvinov2005,Chen2010} in storage rings, or TOF-ICR \cite{Blaum2013}) or PI-ICR \cite{Eliseev2013} in Penning traps, do not offer these four characteristics simultaneously. Typically, they are either very accurate or fast, but not both at the same time. Therefore, MR-TOF-MS is the technique of choice for highly accurate mass measurements of the most exotic nuclides, especially when dealing with short half-lives, low rates, a high amount of contaminants, or low-lying isomers.

Multiple-reflection time-of-flight mass spectrometers (MR-TOF-MS) have been developed for mass measurements at different rare isotope beam (RIB) facilities world-wide \cite{Dickel2015b,Wolf2013b,Schur2013,Jesch2015}. Such MR-TOF-MS measurements have been performed at the forefront in the field \cite{Wienholtz2013,Leistenschneider2018,ito2018,Reiter2018b}. In these measurements, typical mass resolving powers of 100,000 to 200,000 and relative accuracies in the range from $3\times 10^{-7}$ to $10^{-6}$ corresponding to absolute accuracies of 30~keV/c$^2$ to 200~keV/c$^2$. Recently it has even been shown that the uncertainties in MR-TOF-MS mass measurements can be reduced by an order of magnitude, well below $10^{-7}$ \cite{Dickel2015b,Schur2013,Kimura2018} and thus it has reached an accuracy region that was previously accessible with Penning traps only. However, present MR-TOF-MS with a mass resolving power of 100,000 require more than 1000 detected ions to reach a mass accuracy of $10^{-7}$. For rare exotic nuclei such a measurement requires many hours or even days of accelerator beam time. With our superior mass resolving power compared to other MR-TOF-MS we can resolve low-lying isomeric states for the first time\cite{Kimura2018}.

The MR-TOF-MS \cite{Dickel2015b} developed for the FRS Ion Catcher (FRS-IC) \cite{Plass2013} and the MATS experiment \cite{Rodri2010} at FAIR has been designed to overcome these problems \cite{Plass2013b} and to enable mass measurements on the accuracy level of $1\times 10^{-7}$ with a few tens of detected ions. This can be achieved due to its much higher mass resolving power than that of MR-TOF-MS installed at other RIB facilities around the world. However, even at very high resolving powers, low-lying isomers can result in overlapping peaks. Therefore, a data-analysis procedure is required, which is suitable for spectra with overlapping peaks and few events only. Special measures within the data-analysis procedure have to be taken to determine accurately the ground state mass, the excitation energy, the isomer-to-ground state ratios, and their respective uncertainties. 

\begin{figure}
\centering
\includegraphics[width=0.9\linewidth]{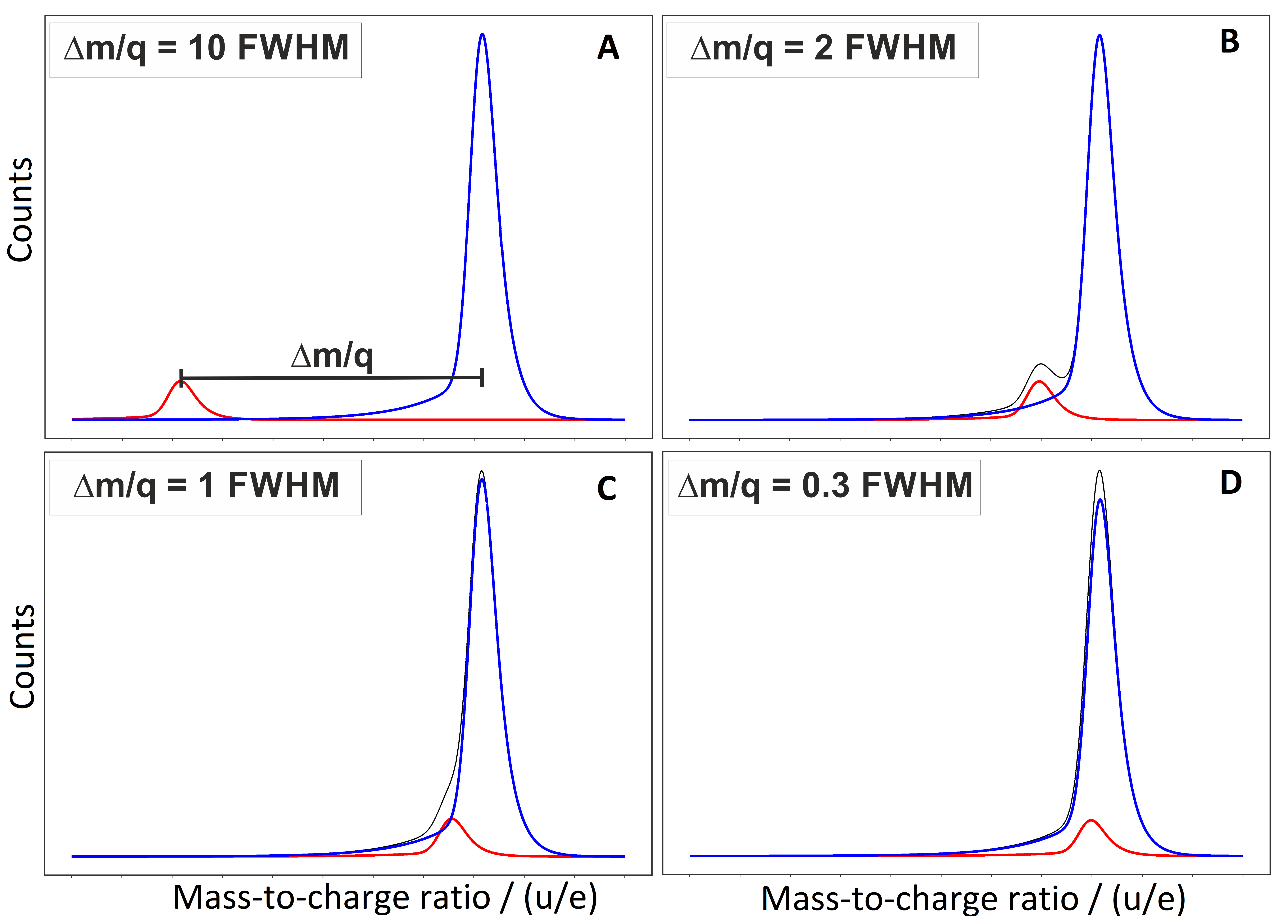}
\caption[Close Lying Peaks Examples]{Schematic mass-to-charge spectra illustrating the four classes of close-lying peaks as defined in the text. The distances between the mass distributions are 10, 2, 1 and 0.3~FWHM for classes A to D, respectively. The peak shape and the abundance ratio of 1 to 10 is selected to be identical for all classes. The black curves show the sum of both distributions.}
\label{fig:Overlap_Examples}
\end{figure}

In this work, close-lying peaks are classified as follows; a corresponding illustration is given in Fig.~\ref{fig:Overlap_Examples}.

\textbf{Class A:} The sum of the distributions reaches almost zero between the peaks. In this case the distributions is considered as non-overlapping and can be analyzed independently.

\textbf{Class B:} The sum of the distributions has a minimum between the peaks, which is significantly larger than zero, even at the minimum. The distributions can be considered as resolved overlapping peaks and an appropriate analysis is performed. In this case the determination of the exact peak shape is important for the extraction of accurate values for the masses and their abundance ratios.

\textbf{Class C:} The sum of the distributions does not have a minimum between the peaks. The existence of a double peak can be determined only from a change in the peak shape, e.g.\ a peak shoulder or a peak broadening, as compared to an individual peak. In this challenging case, the proper determination of the peak shape is a crucial prerequisite in order to first, detect the existence of a second peak and second, to extract the mass values and abundance ratios of the peaks.

\textbf{Class D:} Both peaks overlap almost completely. No change in the peak shape can be observed in the sum of the distributions. In this case the peaks must be considered as unresolved, and the (possible) existence of overlapping peaks can only be taken into account by increasing the uncertainty of the mass value(s).

The mass-to-charge ratio difference between the distributions, the peak shapes and the abundance ratio determine to which class a measurement belongs. While simple data analysis methods can obtain accurate mass and abundance values only for Classes A and B, particularly for peaks with a few events, the data analysis method developed in this work is capable of extracting accurate mass and abundance values even for Class C. For the example shown in Fig.~\ref{fig:Overlap_Examples} this corresponds to an increase in the effective mass resolving power for overlapping peaks by a factor of about three. 

In this publication, mass measurements of $^{238}$U and $^{124}$Xe projectile fragments and of $^{238}$U fission fragments with the MR-TOF-MS of the FRS Ion Catcher at GSI (Germany), the data-analysis procedure and the results are presented. The work includes the measurement with a mass-to-charge ratio difference corresponding down to 280~keV/(c$^2$e) and with down to 11 events. Further details can be found in \cite{Ebert2016,Ayet2018,Hornung2018}.

\section{Experiments} \label{sc_exp}

\begin{figure}
\centering
\includegraphics[width=0.9\linewidth]{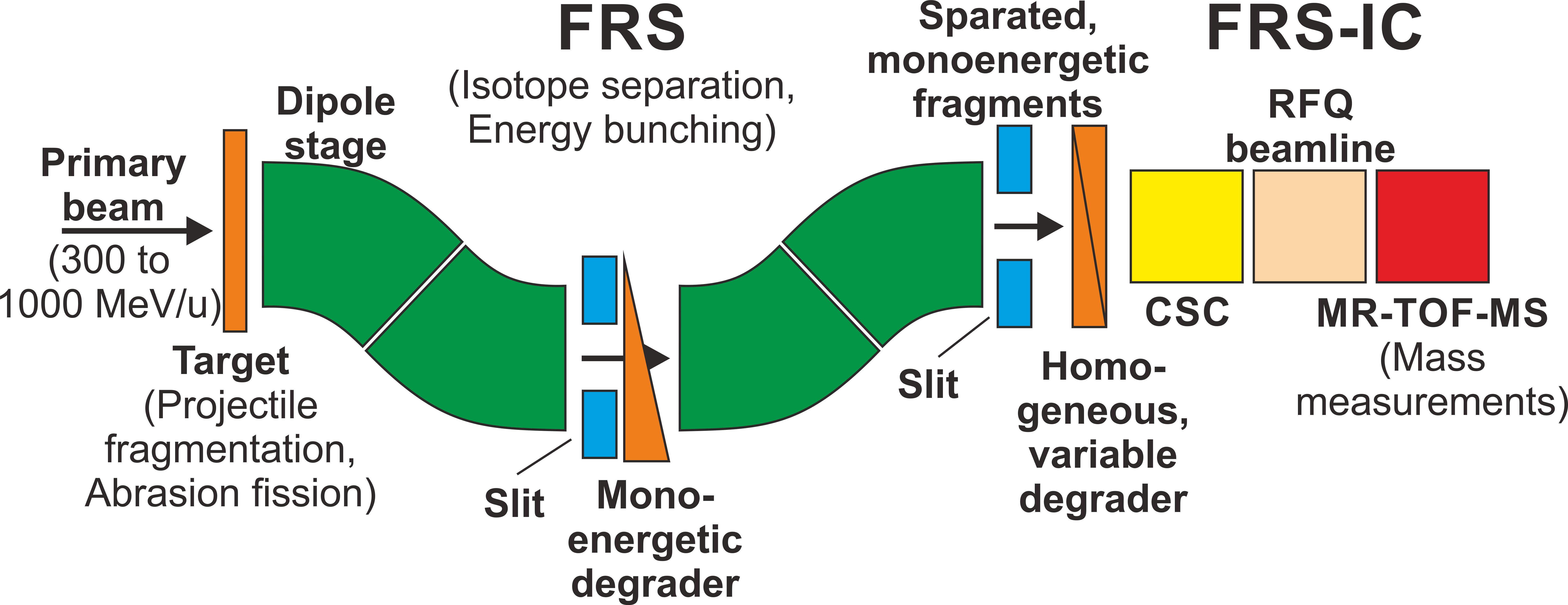}
\caption[Schematic drawing of the experimental setup]{Schematic figure of the experimental setup, including the FRS and the FRS Ion Catcher, which consists of the cryogenic stopping cell (CSC), the RFQ beamline and the MR-TOF-MS.}
\label{fig:FRS-IC_schema}
\end{figure}

The FRS Ion Catcher is an experimental setup installed at the final focal plane of the fragment separator FRS \cite{Geissel1992b} at GSI. The FRS in combination with the FRS-IC enables experiments with thermalized exotic nuclei. In Fig.~\ref{fig:FRS-IC_schema} a schematic view of the FRS-IC with its three main parts is shown: (i) the gas-filled Cryogenic Stopping Cell (CSC) \cite{Ranjan2011,Purushothaman2013,Ranjan2015,Reiter2015} for complete slowing-down of the exotic nuclei produced at relativistic energies, (ii) a beamline, based on Radio Frequency Quadrupoles (RFQ) \cite{Reiter2015,Reiter2011,Miskun2015,Haettner2018} for mass-selective transport and differential pumping. Furthermore, it is equipped with detectors (channeltrons and silicon detectors) for ion counting and $\alpha$-decay spectroscopy and with ion sources for diagnostic purposes. (iii) The MR-TOF-MS \cite{Dickel2015b,Plass2008,Dickel2010} for performing direct mass measurements.

The MR-TOF-MS includes a buffer \mbox{gas-filled} RFQ-based switchyard \cite{Greiner2013,Plass2015,Ebert2016}, which is capable of merging, guiding and splitting low energy ion beams. A combination of a thermal ion source (HeatwaveLabs, Watsonville, CA, USA; mixed source: Ca, Sr and Ba) and an electron-impact ion source is mounted on the top of the switchyard, which generates calibrant ions over a broad mass-to-charge range \cite{Ebert2016}, using various gases such as SF$_6$, Xe or C$_3$F$_8$. Additional ions for calibration were provided by an $^{223}$Ra open $\alpha$-recoil ion source mounted inside the CSC. After the first two experiments the
$^{223}$Ra ion source was replaced by the longer-lived $\alpha$-emitter $^{228}$Th \cite{Rink2017}.

\begin{figure}[htp]
\centering
\includegraphics[width=0.8\linewidth]{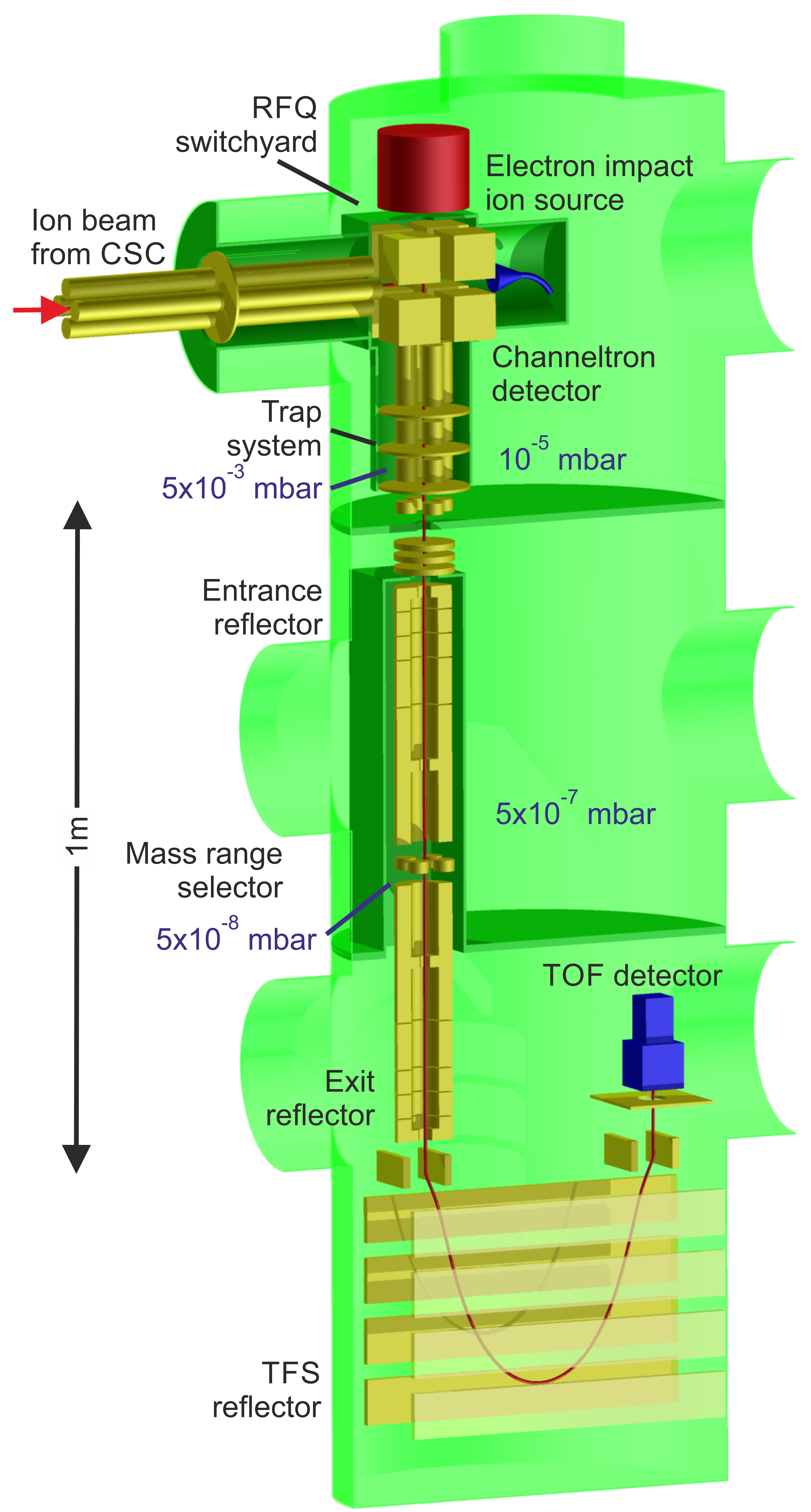}
\caption[MR-TOF-MS schematic view]{Schematic figure of the multiple-reflection time-of-flight mass spectrometer. The ions enter from the CSC in gas-filled radio frequency quadrupoles. They are mixed with ions from the calibration sources and are cooled and bunched in a systems of linear Paul traps. From here, they are injected in the TOF analyzer.}
\label{fig:MR-TOF-MS_schema}
\end{figure}

A schematic view of the MR-TOF-MS of the FRS-IC is shown in Fig. \ref{fig:MR-TOF-MS_schema}. The ions enter the MR-TOF-MS with a kinetic energy of a few eV. In a trap system, they are bunched, their reference potential is raised from about -100~V to 1300~V and they are injected with a repetition rate of 50~Hz towards the analyzer, which is formed by two electrostatic reflectors. The drift tube is at ground potential. The outer electrodes of both reflectors are switched for injection and ejection of the ions towards the time-of-flight (TOF) detector (ETP MagneTOF). In the center of the analyzer is a mass-range selector (MRS) based on a pulsed quadrupolar deflector, which controls the transmitted mass-to-charge window, i.e., ensures an unambiguous mass-to-charge ratio spectrum.

After the ions are ejected from the analyzer, they pass through the time-focus shift (TFS) reflector \cite{Plass2008,Yavor2015,Dickel2017}, which was previously referred to as post-anaylzer reflector according to its position in the device, and impinge on the TOF detector. As an alternative to measurements with the TOF detector, the ions can also be spatially separated with a Bradbury-Nielsen Gate, such that the MR-TOF-MS can be used as an isobar and isomer separator \cite{Plass2008,Dickel2015}. The data acquisition system is based on a commercial Time-to-Digital Converter (TDC), model Ortec-9353. The control of the different electrode potentials along the ion path, are performed via an FPGA-based system \cite{Jesch2016}.

\begin{figure}
\centering
\includegraphics[width=0.9\linewidth]{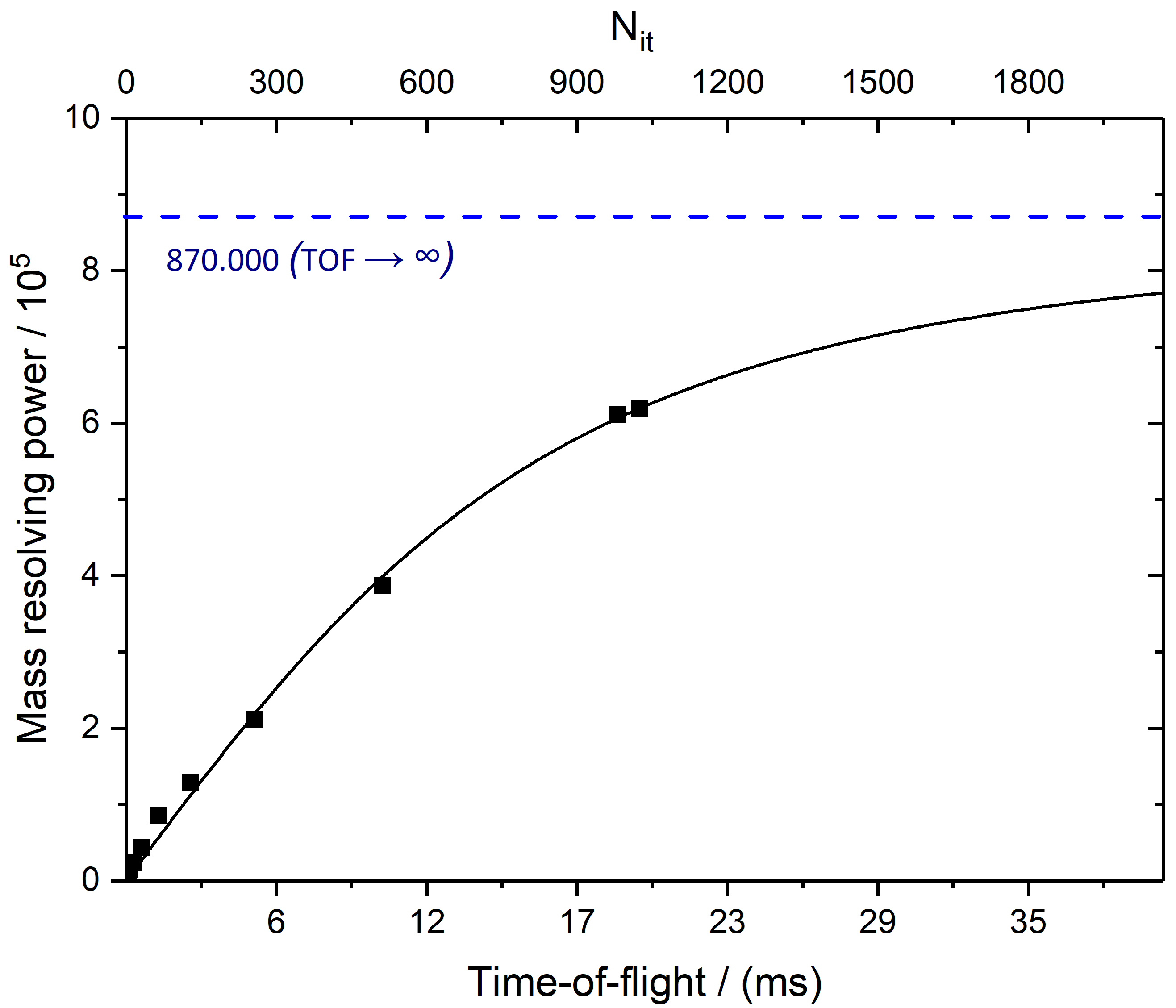}
\caption[MRP Obtained for $^{39}$K]{Mass resolving power obtained with the MR-TOF-MS for $^{39}$K$^{1+}$ ions as a function of the time-of-flight and of the number of turns. The kinetic energy of the ions in the drift tubes was 1,300~eV and the repetition rate was 50~Hz. The measured data are shown as full squares. The solid line represents a fit of Eq.~(\ref{eq:advanced_MRP}) to the data. The dashed line indicates the asymptote obtained from the fit for long flight times. The maximum mass resolving power (FWHM) achieved is 620,000; the asymptote is 870,000. Note that the flight time was limited to 20~ms due to the repetition frequency of the experiment.}
\label{fig:MRP_K}
\end{figure}

Recently, major improvements have been made to the MR-TOF-MS. (i) The kinetic energy of the ions in the drift tube has been increased by a factor 1.7 \cite{Plass2015}. Together with an improved ion-optical tuning this has lead to an increase in the mass resolving power (FWHM) to above 600,000. Figure~\ref{fig:MRP_K} shows the mass resolving power determined in a measurement of $^{39}$K$^{1+}$ ions as a function of the number of turns in the analyzer and the time-of-flight. After a flight time of 2~ms a mass resolving power of more than 100,00 is obtained. After a flight time of 20~ms, i.e.\ the maximum time that is possible for the chosen cycle frequency of 50~Hz, a mass resolving power of 620,000 has been achieved. The asymptote, which is determined by the ion-optical aberration limit, amounts to almost 900,000. (ii) The repetition rate of the MR-TOF-MS has been increased to more than 1 kHz. This increases the rate capability of the device and gives access to shorter-lived nuclei. (iii) The operational reliability and stability have been improved. (iv) The temperature coefficient of MR-TOF-MS has been reduced to 8~ppm/K. (iv) The cleanliness of the buffer gas in the RFQ and trap system has been improved, reducing possible ion losses in the device due to charge-exchange and molecule formation.

In four experiments with the FRS-IC, various exotic nuclei were produced via projectile fragmentation  and abrasion-fission. The primary beams were 300~MeV/u and 1000~MeV/u $^{238}$U ions and a 600~MeV/u $^{124}$Xe ions. The priority of these experiments was the commissioning and characterization of the CSC \cite{Reiter2016}. Details of the experimental conditions and the setups are discussed in the following.

\subsection*{Experiment I: 1000 MeV/u uranium fragmentation}
A 1000~MeV/u $^{238}$U projectile beam was provided from the heavy-ion synchrotron SIS-18 \cite{Blasche1986} with an intensity of up to $7\cdot10^8$ ions per spill, with a spill length of 2~s. A beryllium production target with an areal density of 1.629~g/cm$^2$ with a niobium backing of 0.233~g/cm$^2$ was used at the entrance of the FRS. The ions were energy-bunched via a mono-energetic degrader with an areal density of 4.063~g/cm$^2$ at the central focal plane of the FRS \cite{Geissel1989,Weick2000}. The ions were injected into the CSC with a helium areal density of 3.5~mg/cm$^2$, corresponding to a pressure of 64~mbar, at a temperature of 88~K. For the measurement of $^{211}$Po ions, the areal density was increased to 5.6~mg/cm$^2$, corresponding to a pressure of 95~mbar at a temperature of 86~K.

\subsection*{Experiment II: 1000 MeV/u uranium fission}
Same conditions as for Experiment I were used, but the areal density of the beryllium production target was 6.333 g/cm$^2$.

\subsection*{Experiment III: 300 MeV/u uranium fragmentation}
A 300~MeV/u $^{238}$U projectile beam was provided from the heavy-ion synchrotron SIS-18 with an intensity of up to $2.5\cdot10^8$ ions per spill, with a typical spill length of 1~s. A beryllium production target with an areal density of 0.270~g/cm$^2$ was used. Due to the low primary beam energy, the material in the beamline was minimized. The mono-energetic degrader at the central focal plane had an areal density of 737.1~mg/cm$^2$. The CSC had an areal density of 3.8~mg/cm$^2$ helium, corresponding to a pressure of 75~mbar at a temperature of 99~K.

\subsection*{Experiment IV: 600 MeV/u xenon fragmentation}
A 600~MeV/u $^{124}$Xe projectile  beam was provided from the SIS-18 with an intensity of up to $1\cdot10^9$ ions per spill, with a typical spill length of 500~ms. A beryllium production target with an areal density of 1.622~g/cm$^2$ was used. The CSC had an areal density of 3.8~mg/cm$^2$ helium, corresponding to a pressure of 75~mbar at a temperature of 99~K. 

The priority in Experiments III and IV were high transmission from the production target to the MR-TOF-MS, whereas Experiments I and II were optimized for spatial isotope separation with the FRS. Due to this difference, the abundance ratio of background to ion of interest (IOI) delivered from the FRS was about 1000 times higher in Experiments III and IV with respect to Experiments I and II. In addition, most of the measurements in Experiments III and IV were done with a broader mass-to-charge range, i.e. several mass-to-charge units simultaneously in the spectrum. This increased the background. The amount of background from molecular ions was reduced in Experiment IV by consecutive ion isolation in the RF mass filter of the RFQ beam line, collision-induced dissociation in an RFQ, \cite{Mcluckey1992, Schury2006}) and again ion isolation, also referred to as the Isolation-Dissociation-Isolation (IDI) method \cite{Greiner2017}.

\section{Basics of MR-TOF-MS} \label{sec:Theory}

The time-of-flight (TOF) $t_\mathrm{total}$ in the MR-TOF-MS is the sum of the TOF from the injection trap to the detector without reflections in the analyzer $t_\mathrm{tfs}$, and the TOF for $N_\mathrm{it}$ reflections:

\begin{equation} \label{eq:total_TOF}
t_\mathrm{total} = t_\mathrm{tfs} + N_\mathrm{it} t_\mathrm{it}  ,
\end{equation}
where $N_\mathrm{it}$ is the number of turns in the analyzer and $t_\mathrm{it}$ the TOF for each turn. Similarly, the total flight path $l_\mathrm{total}$ is given by

\begin{equation} \label{eq:total_length}
l_\mathrm{total} = l_\mathrm{tfs} + N_\mathrm{it} l_\mathrm{it}
\end{equation}

where $l_\mathrm{tfs}$ is the path length from the injection trap to the detector and $l_\mathrm{it}$ is the path length for one turn in the analyzer. The ion motion from the injection trap to the detector is made isochronous by the shift of the time focus (time-focus-shift, TFS) by means of the TFS-reflector.  Each turn in the analyzer preserves the isochronicity \cite{Dickel2017}.

In a time-of-flight mass spectrometer the classical relationship between TOF and mass-to-charge ratio is given by

\begin{equation} \label{eq:classical_time_to_mass}
\frac{m}{q} = \frac{2U_\mathrm{eff}t_\mathrm{total}^2}{l_\mathrm{total}^2} , 
\end{equation}

where $m$ and $q$ are the mass and charge of the ion, respectively. $U_\mathrm{eff}$ is the effective voltage, which takes into account the variation of the electric potential along the flight path. The mass resolving power of a time-of-flight mass spectrometer is given by:

\begin{equation} \label{eq:MassResolvingPower}
\left(\frac{m/q}{\Delta (m/q)}\right) = \frac{t_\mathrm{total}}{2 \Delta t_\mathrm{total}}  ,
\end{equation}

where $\Delta t_\mathrm{total}$ is the spread in time-of-flight. In an experiment, the measured time $t_\mathrm{exp}$ includes an time delay $t_{0}$ between the start signal and the real start of ions, caused by the cables and electronic modules, thus:

\begin{equation} \label{eq:measured_TOF}
t_\mathrm{exp} = t_\mathrm{total} + t_\mathrm{0} \;.
\end{equation}

Substitution of Eqs.~(\ref{eq:total_TOF}), (\ref{eq:total_length}) and (\ref{eq:measured_TOF}) into Eq.~(\ref{eq:classical_time_to_mass}) yields:

\begin{equation} \label{eq:time_to_mass_bc}
\frac{m}{q} = \frac{c \left( t_\mathrm{exp}- t_0\right)^2 }{\left(1 + N_\mathrm{it} b\right)^2 } ,
\end{equation}

where $b=l_\mathrm{it}/l_\mathrm{tfs}$ and $c=2U_\mathrm{eff}/l_\mathrm{tfs}^2$.

For the MR-TOF-MS, Eq.~(\ref{eq:MassResolvingPower}) can be written as \cite{Dickel2015b,Dickel2017b}

\begin{equation} \label{eq:ResolutionTurns}
\frac{m/q}{\Delta (m/q)}= \frac{t_\mathrm{tfs} + N_\mathrm{it} t_\mathrm{it}}{2\sqrt{\Delta t_\mathrm{ta}^2 + \Delta t_\mathrm{tfs}^2 + (N_\mathrm{it} \Delta t_\mathrm{it})^2}}  
\end{equation}

where $\Delta t_\mathrm{ta}$ is the turn-around time \cite{Wiley1955}. $\Delta t_\mathrm{tfs}$ is the time spread due to ion-optical aberrations from the injection trap to the detector without reflections in the analyzer. $\Delta t_\mathrm{ta}$ and $\Delta t_\mathrm{tfs}$ together represent the error of $t_\mathrm{tfs}$. $\Delta t_\mathrm{it}$ is the time spread per turn in the analyzer, which is typically dominated by ion-optical aberrations. Dividing by $t_\mathrm{it}$ allows to use measured flight time ratios of a reference ions. To simplify further $t_\mathrm{it}/ (2 \Delta t_\mathrm{it})$ is replaced by $R_\infty$, the mass resolving power for $N_\mathrm{it}=\infty$. The turn-around time is calculated for the reference ion and IOI. The mass resolving power for all mass-to-charge ratios and number of turns measured under the identical ion-optical conditions can be calculated by:

\begin{widetext}
\begin{equation} \label{eq:advanced_MRP}
\left(\frac{m}{\Delta m}\right)\left(q,N_\mathrm{it}\right) = \frac{\frac{t_\mathrm{tfs,ref}}{t_\mathrm{it,ref}}+ N_\mathrm{it}}{2\sqrt{\frac{\displaystyle q_\mathrm{ref}}{\displaystyle q}\left(\frac{\Delta t_\mathrm{ta,ref}}{t_\mathrm{it,ref}}\right)^2 + \left(\frac{\Delta t_\mathrm{tfs,ref}}{t_\mathrm{it,ref}}\right)^2 + \left(\frac{2 N_\mathrm{it}}{R_\infty}\right)^2}} \;.
\end{equation}
\end{widetext}

\section{The data-analysis procedure}
\label{sect:DE}
The analysis procedure of the MR-TOF-MS data has specific requirements and challenges. The peak-fitting routine must be able to cope with overlapping peaks with very low number of events, where the masses of the nuclei, their abundance and their uncertainties have to be determined with the highest accuracy possible. The knowledge of the individual uncertainty contributions are important to obtain higher accuracies in future experiments.  

\begin{figure}[htp]
\centering
\includegraphics[width=0.9\linewidth]{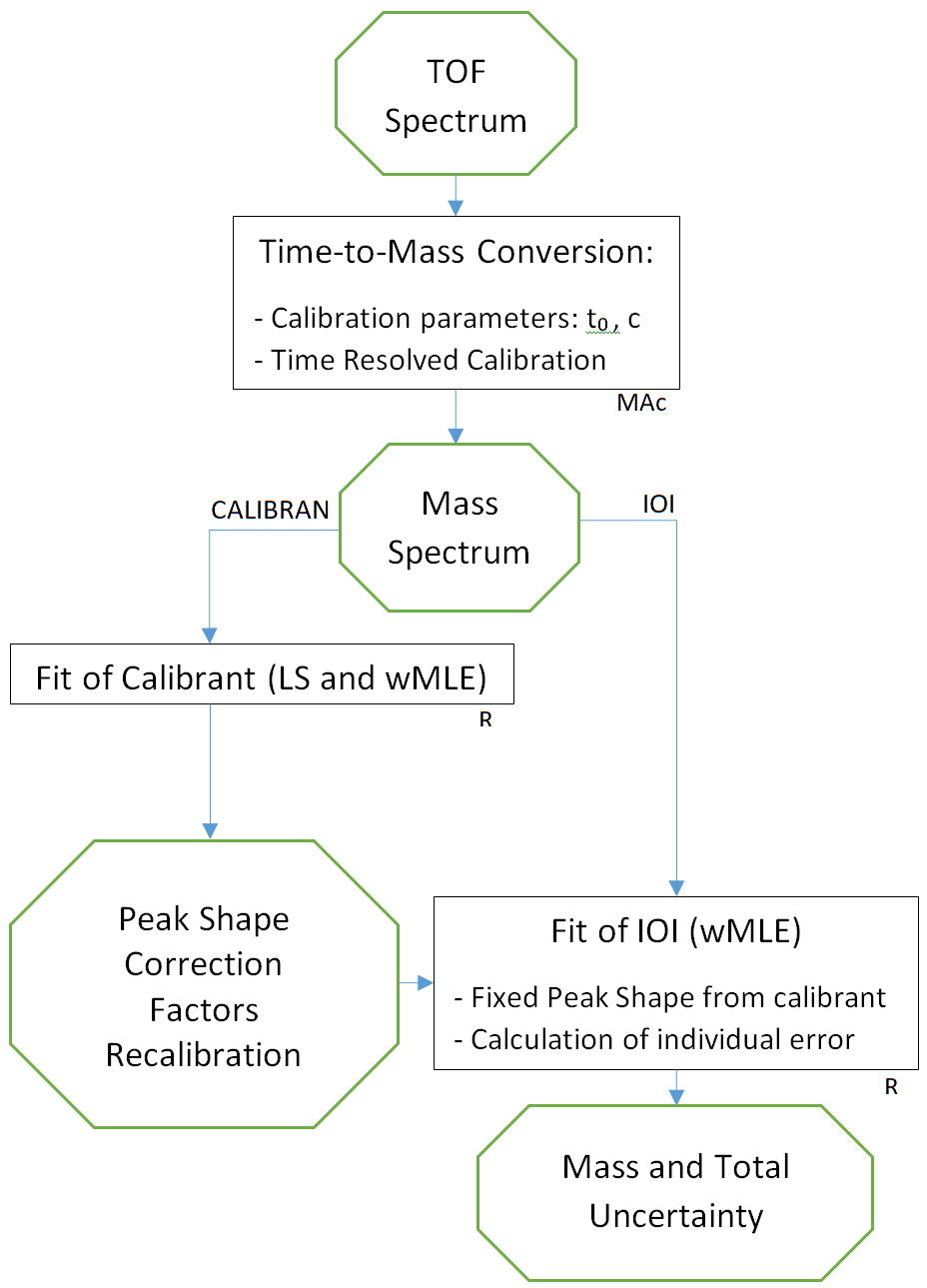}
\caption[data-analysis procedure]{The flow diagram of the data-analysis procedure. Only the main steps are shown. LS is a Least Square fit to determine the peak shape and wMLE is the weighted Maximum Likelihood Estimation of the peak position and area.}
\label{fig:Data_Evaluation_Diagram}
\end{figure}

The TOF of the different nuclei is recorded using the Mass Acquisition (MAc) software \cite{Bergmann2018}. This software is also used for the first steps of the data-analysis procedure. The final analysis is performed in the programming language \textit{R} \cite{R}. In Fig.~\ref{fig:Data_Evaluation_Diagram}, a flow diagram of the data-analysis procedure is shown.  

In the following, the data-analysis procedure well suited for close-lying peaks is described. 

\subsection{The time-resolved mass calibration (TRC)} \label{sec:TRC}
The first step in the data-analysis procedure is the conversion of the TOF spectrum  into a mass-to-charge spectrum. The TOF of a certain mass-to-charge ratio can fluctuate during the measurement, mainly due to two reasons: changes in the potentials applied to the reflector electrodes in the analyzer, and thermal expansion of the analyzer. High-frequency fluctuations in the kHz range and higher are not relevant during the flight of the ions, since the ions are stored in the analyzer for several milliseconds. The fluctuations at lower frequency (down to 0.1~Hz) are minimized by custom made RC low-pass filters. To cope with even slower changes (mHz and lower) a drift correction in the TOF spectrum \cite{Wolf2013b, ito2013} or a time-resolved mass calibration (TRC) can be performed, as it is done in the present work. In TRC the correction is performed in the mass-to-charge spectrum rather than in the TOF spectrum. The TRC is the more powerful method, since it does not introduce additional uncertainties, see \cite{schury2018}. Furthermore, it requires only a single calibrant, even for ions with different number of turns. Note, a relatively coarse mass-to-charge determination is sufficient to perform the TRC. 

Equation~(\ref{eq:time_to_mass_bc}) relates the TOF of an ion to its mass-to-charge ratio. The parameters $c$, $t_\mathrm{0}$ and $b$ have to be determined using calibrant ions. The parameter $t_{0}$ is calculated, before or after the actual mass measurement of the IOI, using the TOF of at least two calibrant ion species that were measured without isochronous turns, in the so-called time-focus shifting (TFS) mode  \cite{Dickel2017}. Two options are possible to determine $b$ and $c$: (i) The high-resolution mass-to-charge spectrum containing the IOI and two calibrant ion species undergoing different turn numbers yield both $c$ and $b$. (ii) The high-resolution spectrum contains calibrants with the same number of turns, then $b$ is calculated from this spectrum and $c$ from the TFS measurement.

The parameter $t_\mathrm{0}$ is constant as long as the electronics and cables are not changed. In a high-resolution measurement with many turns, the TOF during the TFS is short compared to the TOF in the analyzer. Thus temporal drifts during the TFS can be neglected compared to the temporal drifts during the flight in the analyzer. Hence $c$ can be assumed to be constant during a measurement, if a time-dependence of $b$ is allowed for, and $c$ changes only when the ion-optical settings of the TFS mode is changed. Therefore it is sufficient to obtain the time-resolved TOF of a single calibrant.

In order to perform the TRC, a certain number of spectra (typically a few seconds) are summed into a single spectrum (calibration block). A determination of the parameter $b$ is performed for these calibration blocks. For a given number of total events the optimum choice of the number of spectra in a calibration block is a compromise between the accuracy of each $b$ and the time resolution of the TRC. A linear interpolation is applied between the calibration blocks.

The TRC fully corrects the drifts, if the calibrant ions and the IOI experience the same electric fields. The FWHM of the IOI is slightly larger, if they do not experience the same electric fields. Since the width of the peak of the IOI must inferred from the measured width of the peak of the calibrant (Section~\ref{sec:PeakShapeParamaters}), the increase in the peak width was measured in the Experiments I-IV. The peak width increase mainly depends on the time between successive calibration blocks and on the average number of counts in each calibration block. An effective time between calibration blocks is used, since the time between calibration blocks is not necessarily constant:

\begin{eqnarray}
t_\mathrm{trc,eff} &=& \frac{2(t_1 - t_\mathrm{begin})^2 + 2(t_\mathrm{end} - t_n)^2}{t_\mathrm{end} - t_\mathrm{begin}} \nonumber \\
& &+ \frac{\sum_{i=2}^{n}(t_i-t_{i-1})^2 }{t_\mathrm{end} - t_\mathrm{begin}} \label{eq:teff}
\end{eqnarray}

where $t_\mathrm{i}$ is the center of each calibration block, $t_\mathrm{begin}$ and $t_\mathrm{end}$ are the begin and end of the measurement of the IOI, respectively, and $n$ is the number of calibration blocks. The increase of the peak width was calculated as the root-mean-square of the relative mass-to-charge deviation between the true mass-to-charge ratio and the value determined by the linear interpolation between TRC blocks. This increase was tabulated for different times between blocks ($t_\mathrm{trc,eff}$) and different number of test ions in each block. For all measurements performed in Experiments I-IV, the increase of the peak width is obtained from these values.

\begin{figure}
\centering
\includegraphics[width=0.9\linewidth]{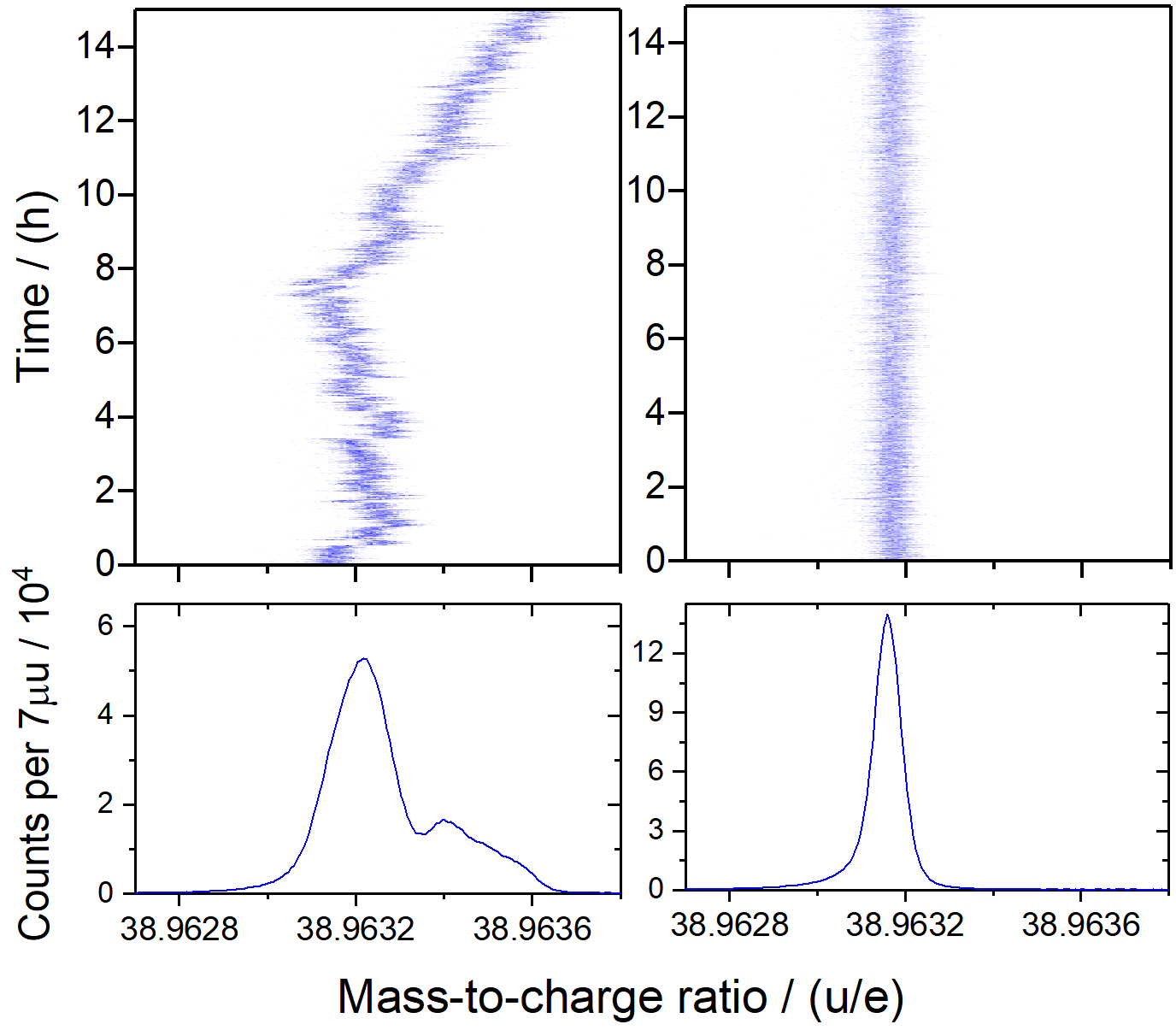}
\caption{Example for the time-resolved calibration (TRC) of a mass measurement of $^{39}$K$^{1+}$ ions that performed 980 isochronous turns in the MR-TOF-MS, corresponding to a total time-of-flight of 18.92~ms. The top diagrams show the peak positions as a function of the experiment time without TRC (left panel) and with TRC (right panel). TRC was performed every 5~s. In the bottom part of the figure the corresponding mass-to-charge spectra are shown. Due to TRC, an increase in the mass resolving power is obtained; the mass resolving power (FWHM) after the TRC amounts to 510,000. Note that all horizontal axes have the same scale for easier comparison.}
\label{fig:TRC_Comp}
\end{figure}

An example for the effect of TRC is shown in Fig.~\ref{fig:TRC_Comp}, where the mass-to-charge spectrum is shown with a single calibration at the beginning of the acquisition and with TRC. 

With the TRC method, the mass-to-charge ratio of each event of the measurement is calculated and saved in list mode for further analysis.

\subsection{Ion Identification}

Based on the mass-to-charge ratio of the ions, an particle identification of the peaks in the mass-to-charge spectra is performed. The identification can be checked according to several criteria: (i) A comparison of absolute and relative detected rates with the results of theoretical simulations or experimental methods, e.g.\ $\alpha$-spectroscopy or particle identification in flight with the detectors of the FRS (ii) A comparison of the identification performed for measurements with two different turn numbers (iii) The coincidence between the events in the MR-TOF-MS and the primary beam (iv) The correlation of the detected events in the MR-TOF-MS with the experimental atomic range of the ions in matter in front of the CSC.

\subsection{Determination of the peak-shape parameters} \label{sec:PeakShapeParamaters}
It is well known that the mass resolving power of TOF mass spectrometers is mass independent, thus peak shapes are the same for all ions with the same charge state. However, their width is proportional the mass of the ion. This has also been studied in detail for MR-TOF-MS in \cite{Jesch2016}. Thus we take the peak shape of a distribution with high statistics and use this for all other distributions measured under the same experimental conditions. In a TOF mass spectrometer the peak shapes can be well described empirically \cite{Purushothaman2017}. They can be determined by fitting a suitable analytical formula to a peak with a large number of events, obtained simultaneously with the IOI. The fit is done by a least-squares (LS) minimization based on the Levenberg-Marquardt algorithm \cite{More1978}. For the LS-fitting the data has to be binned. The Freedmann-Diaconis rule \cite{Freedman1981} is used to determine the bin width $w_\mathrm{bin}$, which is defined as

\begin{equation}
w_\mathrm{bin} = \frac{2 \cdot IQR(x)}{\sqrt[3]{N_\mathrm{counts}}} \; ,
\end{equation}

with $N_\mathrm{counts}$ being the number of events and $IQR$ is the interquartile range ($\pm$~25~\% of the counts around the central event) of a single non-overlapping peak. The analytical formula describing the peaks obtained with the MR-TOF-MS is the Hyper-EMG(L,R) \cite{Purushothaman2017} function, which consists of a weighted sum of a given number of left (L) and right (R) exponentially modified Gaussian (EMG) functions. The parameter of the function that determines the mass-to-charge values is $\mu_\mathrm{G}$, which is the  mean of the Gaussian in the EMG.

\begin{figure}[htp]
\centering
\includegraphics*[width=0.9\linewidth]{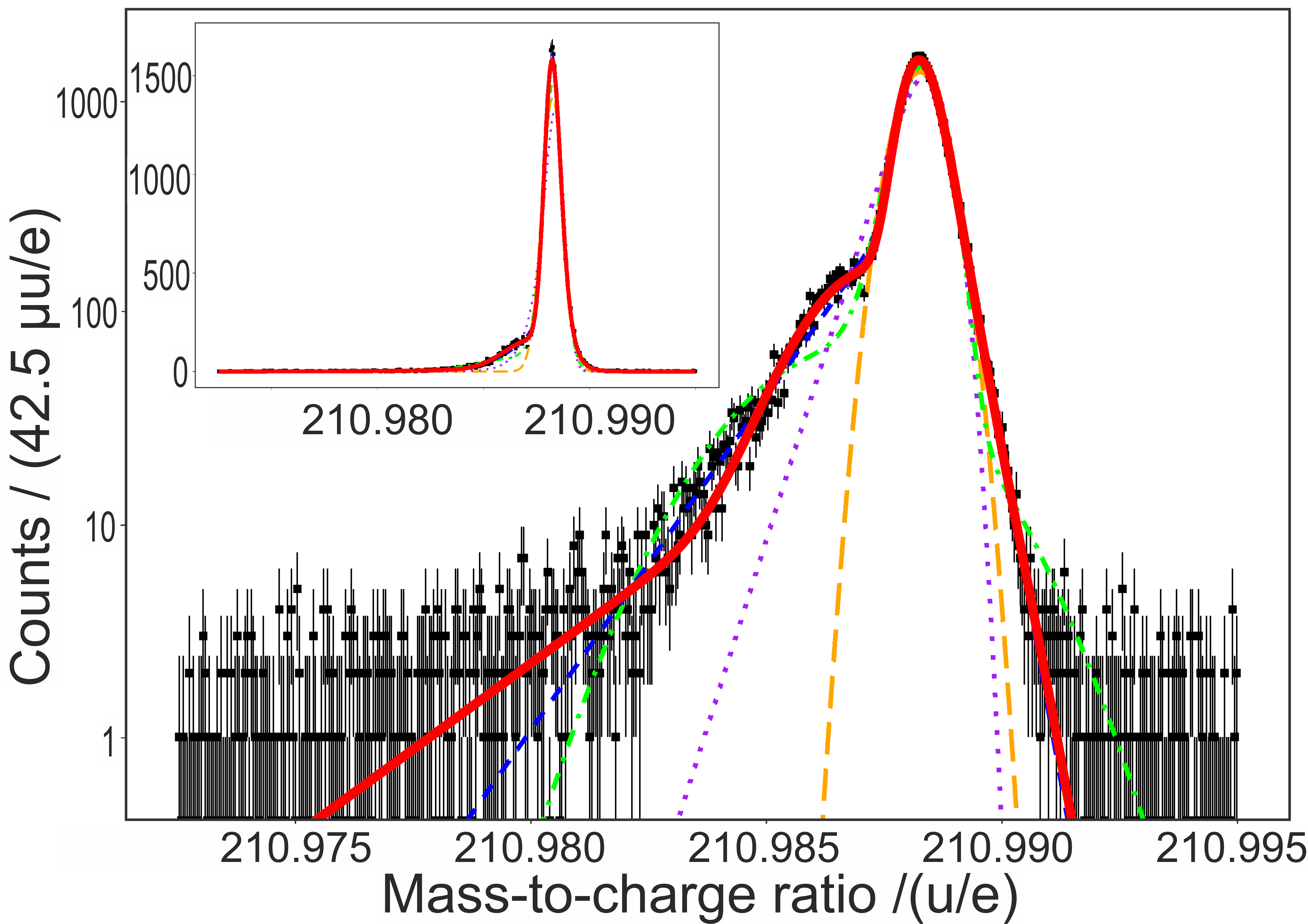}
\caption{Measured mass-to-charge distribution of $^{211}$Pb$^{1+}$ ions, which was used as a calibrant in the mass measurement of $^{213}$Rn ions. The peak shape requires one exponential tail on each side (Hyper-EMG(1,1)) and additionally a Gaussian distribution (red full line) on the left side. The Gaussian distribution results from ion-optical effects. In addition, a Gaussian function (orange long-dashed line) and a Hyper-EMG(1,0) without a Gaussian distribution (purple dotted line), a Hyper-EMG(1,0) plus a Gaussian distribution (green dash-dotted line) and a Hyper-EMG(1,1) (blue dashed line) are shown. The mass-to-charge distribution is shown in linear scale in the inset.  A mass resolving power of 200,000 was achieved for the time-of-flight of 5.8~ms, corresponding to 128 isochronous turns.}
\label{fig:213Rn_211Pb}
\end{figure}

In some cases, a relatively small Gaussian distribution (5-10\% of the area of the main distribution) has to be added in order to get a better determination of the peak shape, see Fig. \ref{fig:213Rn_211Pb}. This Gaussian distribution appears due to ion-optical aberrations in the MR-TOF-MS and its strength depends on the tuning of the analyzer. A uniformly distributed background can be taken into account.

The determination of the peak shape with this high level of accuracy is needed for analyzing data with overlapping peaks. The number of exponentials and the existence of a Gaussian distribution is determined based on the reduced $\chi^2$, the uncertainty of the fit parameters, and the accuracy of the peak-shape model as determined with a Kolmogorow-Smirnow-test (KS-test) \cite{Purushothaman2017}. The peak shape is determined over a mass-to-charge window up to the limits of one event per bin on average.

\begin{figure}[htp]
\centering
\includegraphics*[width=0.9\linewidth]{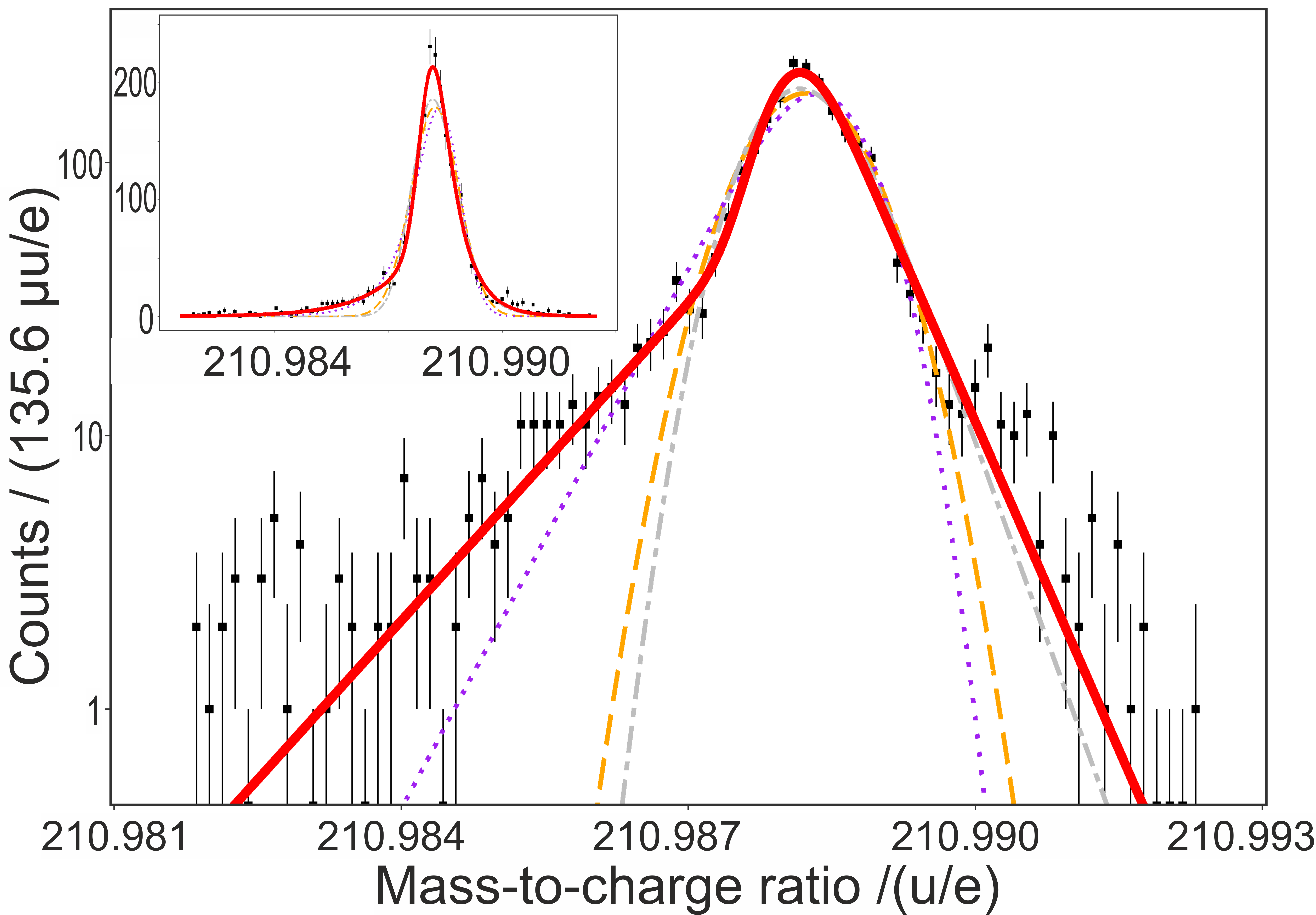}
\caption{Measured mass-to-charge distribution of $^{211}$Pb$^{1+}$ ions, which was used as a calibrant in the mass measurement of $^{212}$At ions. The peak shape requires one exponential tail on each side (Hyper-EMG(1,1)) (red full line). In addition, a Gaussian function (orange long-dashed line), a Hyper-EMG(1,0) (purple dotted line) and a Hyper-EMG(0,1) (gray dashed-dotted line) are shown. The mass-to-charge distribution is shown in linear scale in the inset. A mass resolving power of 180,000 was achieved for the time-of-flight of 5.8~ms, corresponding to 128 isochronous turns.}
\label{fig:212At_211Pb}
\end{figure}

Results of the peak determination obtained with the MR-TOF-MS at the FRS-IC are presented in Figs. \ref{fig:212At_211Pb} and \ref{fig:213Rn_211Pb}. In both figures the data are compared with different Hyper-EMG functions and a regular Gaussian. 

The peak-shape parameters calculated from the calibrant distribution have been scaled to the IOI distribution. The scaling factor takes into account the general increase of the peak width with increasing mass-to-charge ratio, as well as the different contributions to the mass-resolving power discussed in Section~\ref{sec:Theory}. The scaling factor is given by:

\begin{equation} \label{eq:advanced_scaling}
S = \frac{\displaystyle \left(\frac{m}{q}\right)_\mathrm{IOI}\left(\frac{m/q}{\Delta (m/q)}\right)_\mathrm{cal}}{\displaystyle \left(\frac{m}{q}\right)_\mathrm{cal}\left(\frac{m/q}{\Delta(m/q)}\right)_\mathrm{IOI}} ,
\end{equation}

where $(m/q)_\mathrm{IOI, cal}$ are the corresponding mass-to-charge ratio of the IOI and the calibrant. $\left(\frac{m/q}{\Delta (m/q)}\right)_\mathrm{IOI, cal}$ are the corresponding mass resolving power for the IOI and the calibrant ion, as given by Eq.~(\ref{eq:advanced_MRP}). For the scaling of the parameters of the peak tails, the turn-around time is set to zero, because the peak tails are due to optical aberrations. The aberrations do not depend on the turn-around time. The peak broadening, in case IOI and calibrant do not experience the same electric fields, is also taken into account, see Section~\ref{sec:TRC}.

\subsection{Fitting of the mass-to-charge value}  \label{sec:finalmq}
The position of the mass-to-charge value for a given nucleus in the experimental spectrum has been determined by using the well-known weighted maximum-likelihood 
estimation (wMLE) \cite{Hu2002}. However, the measured distribution, was described by the function $f\left(x_i\right)$ determined by the Least Square  (LS) method for the ion used for calibration. 
In the wMLE fit all parameters of the Hyper-EMG, besides $\mu_G$ and the area, are fixed. In the wMLE analysis unbinned data was treated. This has the advantage to avoid additional errors due to the width and position of the bins. In addition, this method is superior to LS fitting for data with low count rates.
The weighting of the data is used to increase the robustness of the fit by minimizing the influence of outliers on the determination of $\mu_G$. The weighting function $w\left(x_i\right)$, dependent on the mass-to-charge value, has been chosen as the truncated natural logarithm of the function $f\left(x_i\right)$. The weighted log-likelihood function is given by:

\begin{eqnarray} \label{eq:weighting}
\mathcal{L} =& \sum_{i=1}^n w\left(x_i\right) \ln{\left[f\left(x_i\right)\right]} \nonumber \\
\text{with}~ w\left(x_i\right) =& 
\begin{cases} 0 , & \text{if}~\ln{\left[f\left(x_i\right)\right]}<0 \\\ln{\left[f\left(x_i\right)\right]} , & \text{otherwise}
\end{cases}
\label{eq:wmle}
\end{eqnarray}
 
After investigations of several types of different weighting functions, by applying bootstrapping, the proposed weighting  function \cite{Ebert2016} was found to be the best compromise between accuracy, outlier suppression and universality for the atomic mass determination.

\subsection{Effect of the mass-range selector (MRS) }\label{mrs_cor}
When measuring ion species, which undergo different numbers of turns, the mass-to-charge spectrum can be ambiguous. By isolation of ions of a certain mass-to-charge range, i.e.\ removal of all ions outside this mass-to-charge range, an unambiguous identification of all ion species can be obtained \cite{Dickel2015b}. Isolation is performed by the mass-range selector (MRS), which is a deflector mounted in the middle of the analyzer. By switching the MRS deflector between transmission and deflection with proper timing only the ions of interested are transmitted.

A turn in the analyzer with the MRS in operation is referred to as an isolation cycle (IC). In each IC the MRS is switched four times. Each switching causes a small change in the voltages applied to the MRS. As a consequence, the flight times of the transmitted ions shift. For a given electrical setup of the MRS, the shift is the same for all ions of the same mass-to-charge ratio and for the same number of IC. The shift increases linearly with number of IC $N_\mathrm{ic}$, and according to Eq.~(\ref{eq:classical_time_to_mass}) the shift in the time-of-flight is proportional to $(m/q)^{1/2}$. Therefore, a measurement of the MRS shift can be made and be used as the basis for a correction of the MRS shift in all subsequent mass measurements. 

The result of a measurement of the MRS shift is shown in Fig.~\ref{fig:MRS_Shift}. The measured shift in the time-of-flight $\Delta t_\mathrm{mrs}$ was normalized with $(m/q)^{1/2}_\mathrm{ref}$ of the ion, here $^{133}$Cs ions, and fitted to the equation:

\begin{figure}
\centering
\includegraphics[width=0.9\linewidth]{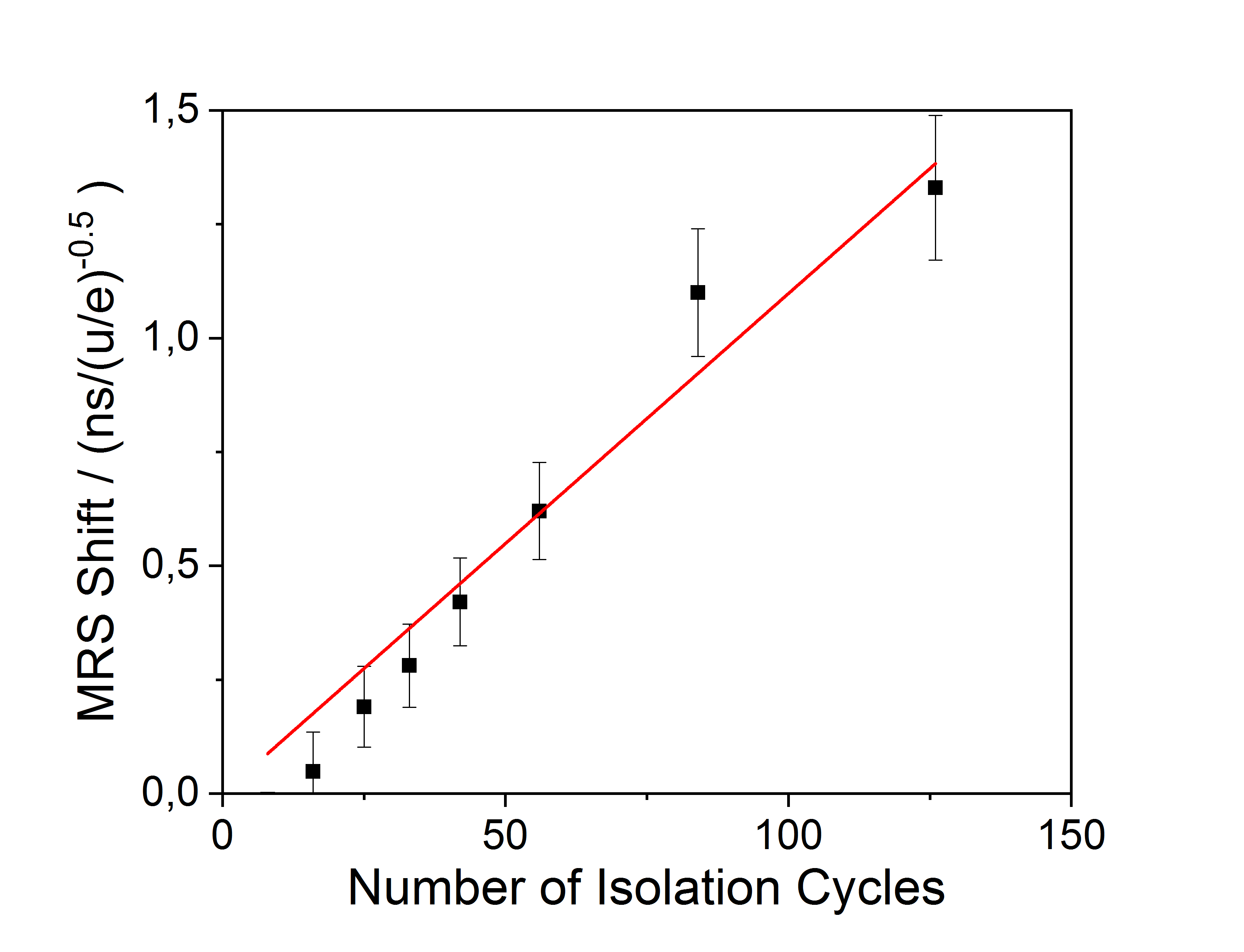}
\caption[MRS Shift]{Measured shift of the time-of-flight of $^{133}$Cs ions due to the MRS as a function of the number of isolation cycles. The shift is normalized by $(m/q)^{1/2}$ (for details see text). The lines shows a fit of Eq.~({\ref{eq:FitMRSShift}}) to the data points. The resulting fit constants are $A_\mathrm{mrs}=1.1 \cdot 10^{-11}$ s /$\sqrt{\mathrm{u/e}}$.}
\label{fig:MRS_Shift}
\end{figure}

\begin{equation} \label{eq:FitMRSShift}
\frac{\Delta t_\mathrm{mrs}}{\displaystyle \sqrt{\left(\frac{m}{q}\right)_\mathrm{ref}}}=A_\mathrm{mrs} N_\mathrm{ic} .
\end{equation}

The MRS shift for all other ions can be calculated from

\begin{equation}
\Delta \left(\frac{m}{q}\right)_\mathrm{mrs} = \frac{2 A_\mathrm{mrs}}{t_\mathrm{total}} \left(\frac{m}{q}\right)^\frac{3}{2} N_\mathrm{ic} r_\mathrm{mrs}    ,
\end{equation}

where Eq.~(\ref{eq:MassResolvingPower}) has been used to convert the shift in the time-of-flight into a corresponding shift in the mass-to-charge ratio. The factor $r_\mathrm{mrs}$ applies to the case where the MRS was in operation for a part of the experiment only. It is the ratio, during which the MRS was in operation, to the overall duration of the measurement. This MRS shift is applied as a correction to the mass-to-charge values of both the IOI and the calibrant ion. 

This method for correcting the MRS shift was tested in a measurement of $^{123}$Xe ions under conditions, where the magnitude of the shift due the MRS was maximized. Without the correction, the measured mass of $^{123}$Xe ions deviated by more than three standard deviations from the literature mass. After applying the MRS shift correction and adding its uncertainty (see Section~\ref{sec:MRSUncertainty}) to the overall uncertainty of the measurement, the measured mass agreed with the literature mass within one standard deviation.

\subsection{Overlapping peaks}\label{overlapping_peaks}

The weighting of the wMLE method causes a smaller distance between the fitted values of overlapping peaks. This effect is stronger for closer peaks and a larger difference in area. A correction algorithm for overlapping peaks of class B and C is used, to cope with this effect\cite{Ebert2016}. In this iterative algorithm first a wMLE fit is performed, which gives the initial mass-to-charge ratio of each peak. From this the distance between peaks is calculated. Then, $N$ spectra are simulated with these parameters and fitted, and an average mass-to-charge distance ($\epsilon$) of the simulated data is calculated. A new simulated data set is generated and fitted with a mass-to-charge distance corrected by $\epsilon$. The procedure is repeated until $\epsilon$ is smaller than the threshold. A final fit with a fixed distance between the peaks, corresponding to the distance of the last simulated spectrum, is performed. 

\begin{figure}[htp]
\centering
\includegraphics[width=0.9\linewidth]{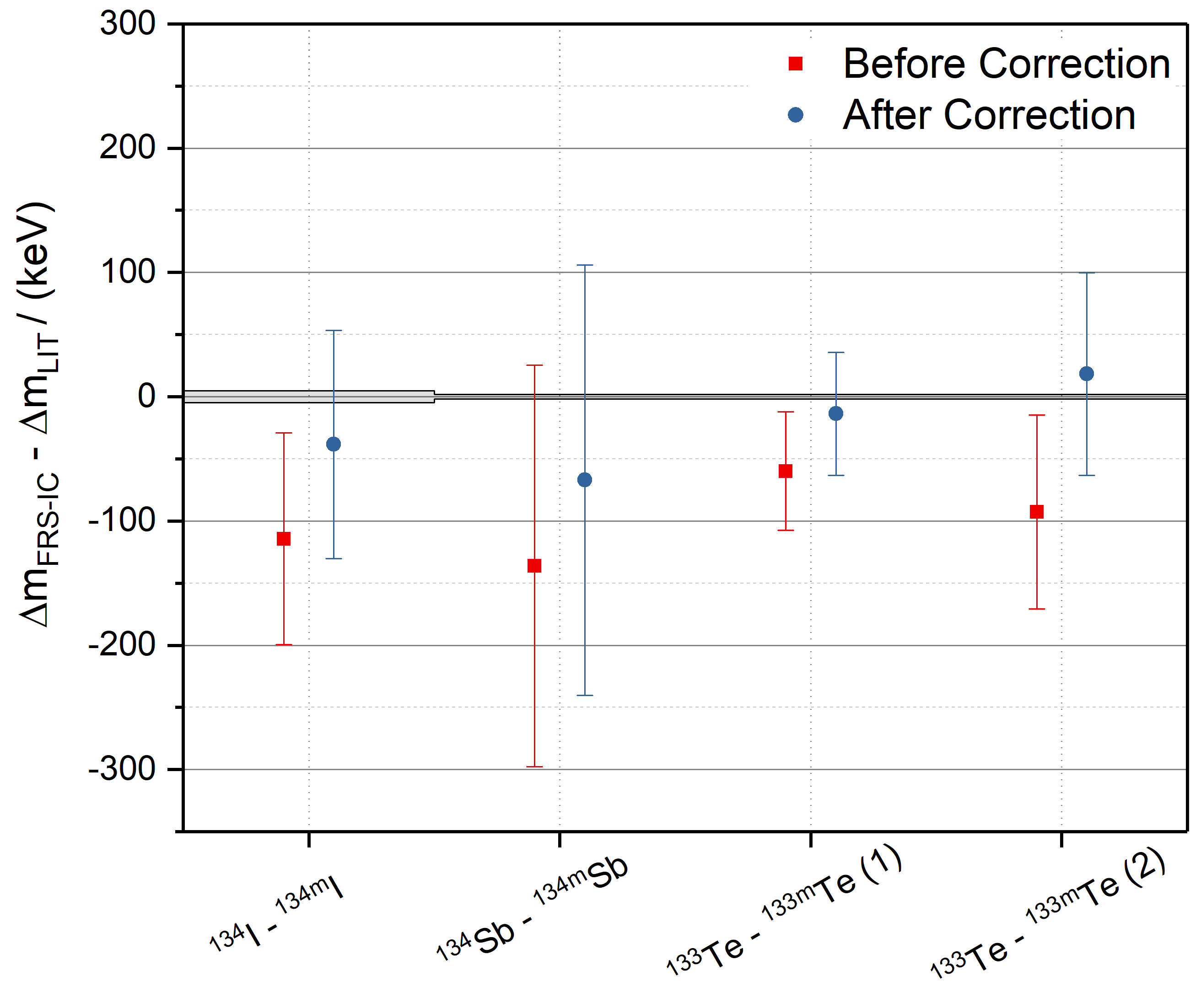}
\caption[Bias Correction Examples]{Example for the effect of the correction algorithm for fitting overlapping peaks of Class C. The deviations of measured and literature values for the mass-to-charge difference of isomeric and ground states determined for different nuclides, are shown without the correction (red data points) and with the correction (blue data points). The literature values were taken from the AME2016 \cite{Wang2017}. The error bars of the measurements before the correction do not include the uncertainty of the correction. The grey band centered around the horizontal axis shows the error bars of the mass-to-charge differences as obtained from literature values.}
\label{fig:Bias_Correction_2Peaks}
\end{figure}

The algorithm for overlapping peaks of Class C was tested with simulated and real data. The simulated data have a similar peak shape as shown in Fig. \ref{fig:FRSIC14_133Te_Full}. The distance between the peaks is $ \Delta m/q = 350 \cdot 10^{-6}$ u. Without correction the peak distance obtained was $ \Delta m/q~=~(322.9\pm~4.9)\cdot~10^{-6}$~u/e, and with correction, $ \Delta m/q~=~(350.3\pm~5.5)\cdot~10^{-6}$~u/e, demonstrating the powerful correction. 
In Fig. \ref{fig:Bias_Correction_2Peaks}, examples of the measured nuclides are presented, where this special algorithm was applied. The deviations from the literature values are presented before and after the correction. All these pairs of ground and isomeric states correspond to overlapping peaks of Class C. For these examples the mass-to-charge difference tends to be underestimated without the correction. After the correction the values are in perfect agreement with the literature values.For overlapping peaks of Class B, this correction is negligible.

\subsection{Relativistic correction}
The kinetic energy of the ions in the MR-TOF-MS was treated classically. Therefore the mass deviation between relativistic and classical treatment is estimated by a Taylor expansion of the relativistic formula:
\begin{equation} 
m = \frac{E_\mathrm{kin}}{c_0^2 (\gamma -1)}\;.
\end{equation} 

The uncertainty of the classical treatment ($\Delta m_\mathrm{relativistic}$) can be estimated by calculating the ratio between the first order correction and the classical part. Higher order corrections are negligible. If the velocity is replaced by the kinetic energy (classical) one obtains
\begin{equation}
	\Delta m_\mathrm{relativistic} = -\frac{6}{4} \cdot \frac{E_\mathrm{kin}}{c_0^2}
\end{equation}
that is a mass-independent correction, which depends only on the average kinetic energy. This energy has been determined to an accuracy of about 1\% by a combination of measurements and simulations. The resulting correction for the mass-to-charge value of all ions is -1.73~keV/$c_0^2$ multiplied by the charge state of the ions. The calibration formula, including relativistic effects, is obtained by adding the correction ($\Delta m_\mathrm{relativistic}$) to Eq. \ref{eq:time_to_mass_bc}. With this, the uncertainty is reduced to a few $eV/c_0^2$ even in the cases of largest possible mass-to-charge differences.

\subsection{Final mass-to-charge value}\label{final_m}

The final mass-to-charge value of the IOI is obtained from the fitted mass-to-charge value of IOI $(m/q)_\mathrm{IOI,wMLE}$, calibrant ion $(m/q)_\mathrm{cal,wMLE}$, and the literature value for the mass-to-charge value of the calibrant ion $(m/q)_\mathrm{cal,lit}$. It is given by

\begin{equation} \label{eq:final_IOI_mass}
\left(\frac{m}{q}\right)_\mathrm{IOI} = \frac{\displaystyle \left(\frac{m}{q}\right)_\mathrm{IOI,wMLE}}{\displaystyle \left(\frac{m}{q}\right)_\mathrm{cal,wMLE}} \left(\frac{m}{q}\right)_\mathrm{cal,lit}\;.
\end{equation}
Note that up to this step in the data-analysis procedure, the mass-to-charge scale has been established using an interpolated median \cite{Ramac2009} in the TRC. Therefore, both the IOI and the calibrant ion have to be fitted with the wMLE to obtain the final mass-to-charge value and corresponding uncertainty.

\subsection{Uncertainty contributions}
The final mass-to-charge uncertainty is calculated by adding in quadrature the various uncertainties described in the following. Effects like the earth magnetic fields are not discussed in detail, because they have been found to be negligible. The uncertainty of mass-to-charge differences of close-lying peaks, e.g., excitation energy of isomers, partially cancel. In this case, the remaining uncertainties are due to the statistics, the unresolved peaks, the overlapping peaks, the space charge and the dead-time.

\subsubsection{Statistical uncertainty}
In samples obeying a normal distribution, the statistical uncertainty of the mean value is given by:

\begin{equation}
\sigma_\mathrm{stat} = \frac{1}{\sqrt{8\ln2}} \frac{\mathrm{FWHM}}{\sqrt{N_\mathrm{counts}}} ,
\end{equation}

where FWHM is the full width half maximum of the normal distribution and $N_\mathrm{counts}$ the number of samples. For the wMLE fit with a Hyper-EMG function, an empirical approach has been taken to determine the statistical uncertainty, because there is no analytical solution for this case.


Random numbers are drawn (same number of events as measured) according to the distribution function determined for the calibrant and IOI and are fitted with the wMLE. This is repeated many times (typically 1000). The standard deviation of the mass-to-charge values obtained from this, is the statistical uncertainty of the calibrant or the IOI. The statistical uncertainties determined in this way for mass measurements of $^{213}$Fr ions in Experiment I as a function of the number of counts per spectrum are shown in Fig.~\ref{fig:statistical_error}. The equation:

\begin{figure}
\centering
\includegraphics*[width=0.9\linewidth]{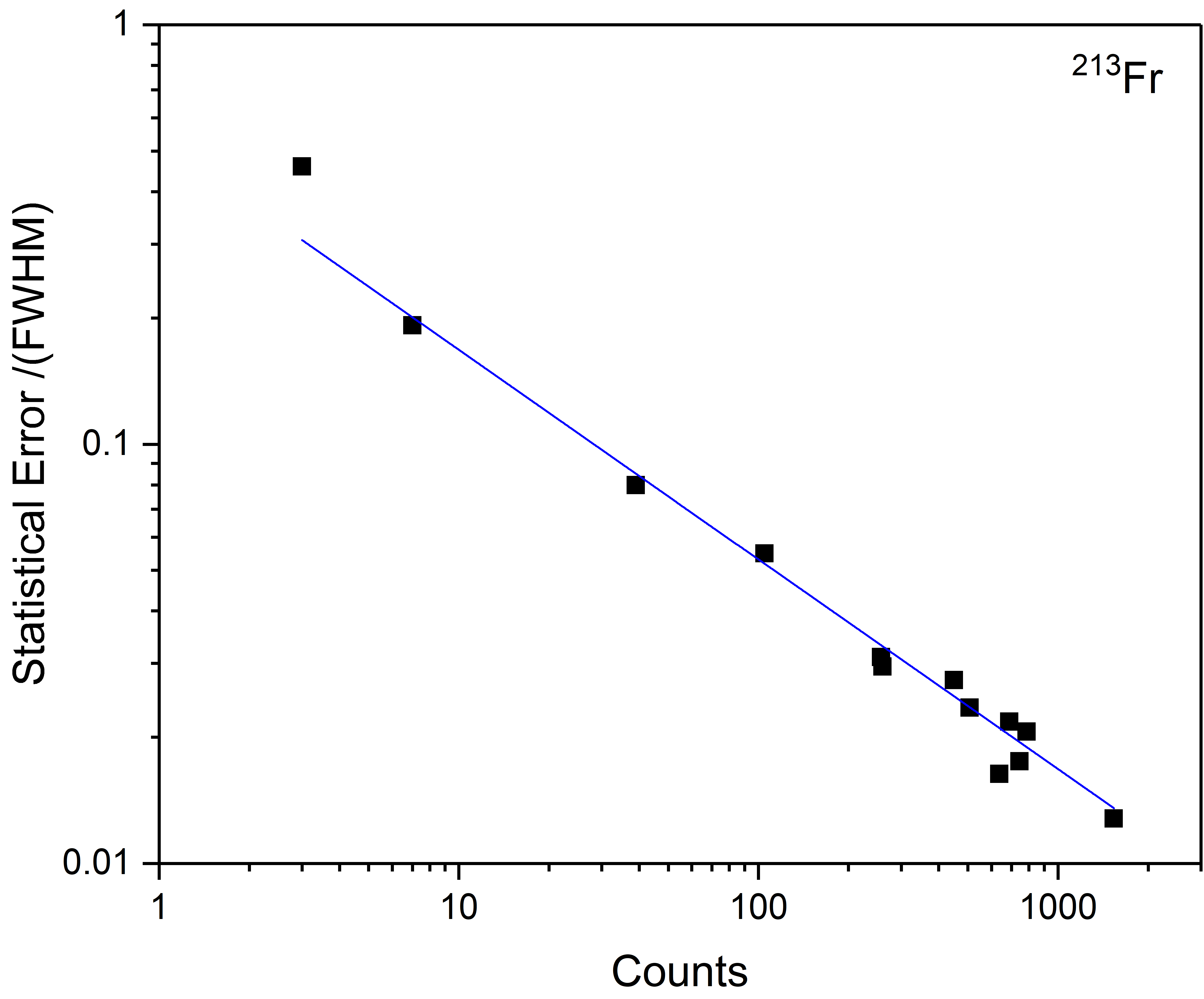}
\caption{Statistical uncertainty normalized to the FWHM of the peak as a function of the number of measured $^{213}$Fr ions per spectrum of Experiment I. The mean value for the FWHM of the distribution in this measurement was $1.03 \cdot 10^{-3}$~u/e.}
\label{fig:statistical_error}
\end{figure}

\begin{equation}
\sigma_\mathrm{stat} = A_\mathrm{stat} \frac{\mathrm{FWHM}}{\sqrt{N_\mathrm{counts}}}
\end{equation}

was fitted to the data points. A value of $A_\mathrm{stat} = (0.53 \pm 0.02)$ was obtained. The value is typical for the measurements presented in the paper.  It is about 25~\% larger than the uncertainty of the mean of a normal distribution with the same FWHM and area. The reason for this difference is the weighting used in the wMLE fit and the differences in the peak tails between normal distribution and Hyper-EMG. 

\subsubsection{Peak-shape uncertainty}
The uncertainty of the determined peak-shape parameters are considered in the following. It is assumed that the uncertainties of the different parameters contribute independently to the uncertainty of the mass-to-charge value ($\mu_\mathrm{G}$). Each parameter of the model is changed by its uncertainty, while keeping the others unchanged, and the peak is fitted again with the wMLE method. The relative change in the obtained mass-to-charge ratio is calculated for the calibrant and the IOI. The deviations are calculated for all parameter and are quadratically added to obtain the peak-shape uncertainty.

\subsubsection{Uncertainty of the time-resolved calibration (TRC)}

 The TRC has an uncertainty in addition to the described peak broadening (Section~\ref{sec:TRC}) of the IOI compared to the calibrant. This is caused by the uncertainty of the interpolated calibration parameter $b$. Likewise for the peak broadening, this effect has been determined for all experiments in dedicated measurements. A measure for this deviation is the square root of the quadratic difference between the RMS of the  mean relative mass-to-charge deviation for each calibration block and the RMS value of the statistical uncertainties of each calibration block. The latter is given by $\mathrm{RMS}/N_\mathrm{trc,eff}^{1/2}$, where $N_\mathrm{trc,eff}$ is the effective number of calibration blocks

\begin{equation} \label{eq:O}
N_\mathrm{trc,eff} = \frac{t_\mathrm{end}-t_\mathrm{begin}}{t_\mathrm{trc,eff}}
\end{equation}

These values were determined for different effective times between blocks $t_\mathrm{trc,eff}$ and different number of counts of the calibrant in each calibration block.

This uncertainty contribution can then be calculated from the determined deviation $A_\mathrm{trc}$ for a given $t_\mathrm{trc,eff}$ and number of calibrant ions as

\begin{equation} \label{eq:TRC_Error}
\sigma_\mathrm{trc} = \frac{A_\mathrm{trc}}{\sqrt{N_\mathrm{trc,eff}}} \left( \frac{m}{q} \right)_\mathrm{ioi}
\end{equation}

\subsubsection{Calibration uncertainty}
This uncertainty includes the statistical uncertainty and the uncertainty of the literature mass of the calibrant. Based on Eq.~(\ref{eq:final_IOI_mass}), the uncertainty due to the calibrant can be written as


\begin{eqnarray} 
\sigma_\mathrm{cal} &=& \left(\frac{m}{q}\right)_\mathrm{IOI} \left[\left(\frac{\Delta\left(\frac{m}{q}\right)_\mathrm{cal,lit}}{\left(\frac{m}{q}\right)_\mathrm{cal,lit}}\right)^2 \right. \nonumber \\
& &+ \left. \left(\frac{\Delta\left(\frac{m}{q}\right)_\mathrm{cal,wMLE}}{\left(\frac{m}{q}\right)_\mathrm{cal,wMLE}}\right)^2 \right]^{1/2} \label{eq:p_calibration_error} 
\end{eqnarray}

where $\Delta (m/q)_\mathrm{cal,lit}$ is the literature uncertainty for the calibrant mass and $\Delta (m/q)_\mathrm{cal,wMLE}$ the statistical uncertainty in the fitted calibrant mass-to-charge ratio. In case the mass value of the calibrant ions in the literature changes $\left(\frac{m}{q}\right)^\mathrm{new,old}_\mathrm{cal}$ a new and updated mass for the ion of interest $\left(\frac{m}{q}\right)^\mathrm{new}_\mathrm{IOI}$can be calculated by the following relation:
\begin{equation} \label{eq:new_lit_mass}
	\left(\frac{m}{q}\right)^\mathrm{new}_\mathrm{IOI} = \left(\frac{m}{q}\right)^\mathrm{old}_\mathrm{IOI} \frac{\left(\frac{m}{q}\right)^\mathrm{new}_\mathrm{cal}}{\left(\frac{m}{q}\right)^\mathrm{old}_\mathrm{cal}} \;,
\end{equation}
where $\left(\frac{m}{q}\right)^\mathrm{old}_\mathrm{IOI}$ is the old mass-to-charge value of the IOI.

\subsubsection{Uncertainty of the calibration parameters ($\Delta t_{0}$, $\Delta c$ and $\Delta b$)}
The uncertainties of the calibration parameters, $t_{0}$ and $c$, $\Delta t_{0}$ and $\Delta c$, are determined during the conversion of the time-of-flight into a mass-to-charge ratio spectrum. The peak positions of the calibrant ions, used for determining these parameters, are shifted in time separately by plus/minus their uncertainty and the calibration parameters are recalculated. The maximum deviation of the calibration parameters for each calibrant species from the overall mean value is calculated and summed quadratically for each calibration parameter separately. The resulting values are used as the individual uncertainties $\Delta t_{0}$ and $\Delta c$. The mass-to-charge uncertainty ($\sigma m_{t_{0}}$) due to the uncertainty in $t_0$ is given by:

\begin{eqnarray}
\sigma m_{t_{0}} &=& \frac{2 \sqrt{c}\left(\frac{m}{q}\right)_{IOI} }{1 +N_{IT} b} \cdot \Delta t_{0} \cdot \left(\frac{1}{\sqrt{\left(\frac{m}{q}\right)_{cal}}} \right. \nonumber \\
& &\left. - \frac{1}{\sqrt{ \left|\left(\frac{m}{q}\right)_{cal}-\left(\frac{m}{q}\right)_{IOI}\right| +\left(\frac{m}{q}\right)_{cal}}}\right) .
\end{eqnarray}

The uncertainty in determining $c$ and $\Delta c$ results in an uncertainty component of the final mass-to-charge value in units of u/e, which is given by
\begin{widetext}
\begin{equation}
\sigma m_\mathrm{c} = \left|\left(\frac{m}{q}\right)_\mathrm{IOI} \left(c\right) - \left(\frac{m}{q}\right)_\mathrm{IOI} \left(c \pm \Delta c \right) \right| = 
\left|\frac{c \left( t_\mathrm{IOI}-t_{0} \right)^2}{\left(1+b N_\mathrm{it,IOI} \right)^2}-\frac{\left(c\pm \Delta c \right) \left( t_\mathrm{IOI}-t_{0} \right)^2}{\left(1 + \frac{N_\mathrm{it,IOI}}{N_\mathrm{it,cal}}\left(\sqrt{\frac{\left(c \pm \Delta c \right) \left( t_\mathrm{cal}-t_\mathrm{TFS}\right)^2}{\left(\frac{m}{q}\right)_\mathrm{cal}}} -1\right) \right)^2}\right|
\end{equation}
\end{widetext}

When the calibrant ion and the IOI have the same number of isochronous turns in the analyzer, the uncertainty component $\Delta m_\mathrm{c}$ is zero. This is because the calibration parameters $c$ and $b$ are not independent and the uncertainty is described fully by the uncertainty of $b$. The higher the difference of turns between the calibrant used for $b$ and the IOI, the higher the contribution of the uncertainty of $c$. The uncertainty of $b$ is included in the calibration uncertainty, \ref{final_m}.

\subsubsection{Scaling-parameter uncertainty}
The uncertainty is calculated from the difference between the scaling factors for same and different resolving powers for calibrant and IOI (see Eq. \ref{eq:advanced_scaling} for scaling factor). This uncertainty is quadratically added to the parameter uncertainty of $\sigma$ and the different $\tau$ before the values are used for the peak-shape uncertainty determination. The scaling uncertainty also influences the statistical error. This uncertainty is estimated by multiplying the statistical error with the normalized mean of the two scaling factors. This uncertainty is smaller than 5\% of the total uncertainty of all cases presented here. 

\subsubsection{Uncertainty due to the Mass-range selector (MRS)} \label{sec:MRSUncertainty}

The uncertainty $\sigma_\mathrm{mrs}$ due to switching fields of the MRS is assumed to be 50\% of the correction in the final mass-to-charge value due to the MRS (Section~\ref{mrs_cor}).

\subsubsection{Uncertainty of the Non-ideal ejection (NIE)}
\begin{figure}[htp]
\centering
\includegraphics[width=0.9\linewidth]{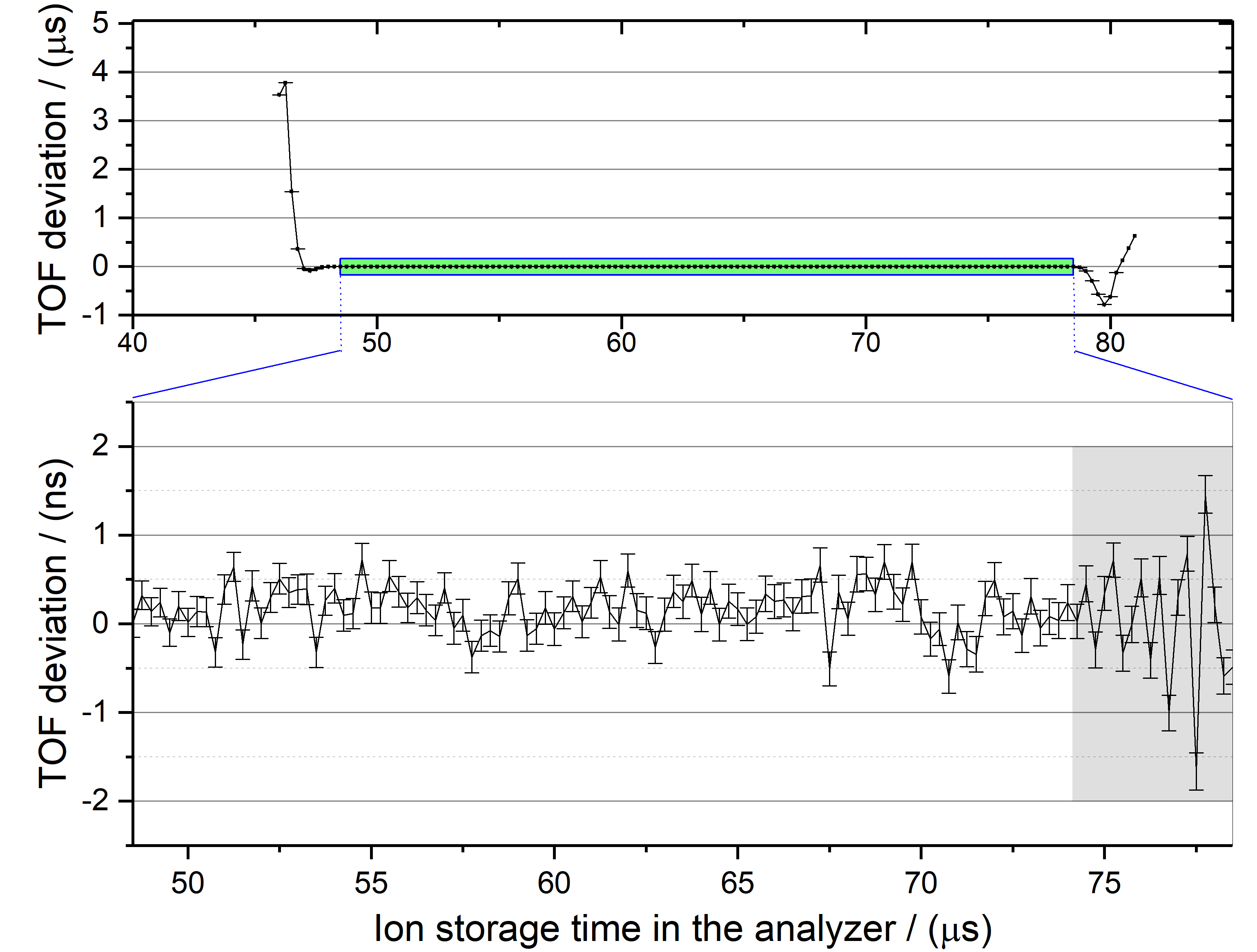}
\caption[NIE Plot 2016]{Measured variation of the TOF of $^{133}$Cs$^{1+}$ ions after two isochronous turns in dependence of the storage time of the ions in the analyzer. The lower panel shows a zoom to the time interval for which a mass determination can be performed. The non-ideal-ejection (NIE) uncertainty is calculated from that interval. For the grey range in the lower panel, an increased NIE uncertainty of 2~ns is used. The TOF deviations are the difference between the TOF for the individual measurement and the mean of all measurements in the lower panel range.}
\label{fig:NIE_Plot_2016}
\end{figure}

The voltages applied to the analyzer electrodes need time, some $\mu $s, to reach the final value. During extraction from the analyzer, ions with different mass-to-charge ratios that are spatially separated might experience different fields. The deviation of the TOF is measured depending on the opening time of the output reflector to determine this influence. Since this effect occurs only during ejection from the analyzer, its absolute value is constant and independent on the number of turns. Therefore, measurements with a low number of turns (typically 2) are used to quantify it. The data taken with $^{133}$Cs ions (two turns) for the Experiments III and IV is shown in Fig. \ref{fig:NIE_Plot_2016}. The ions in the ``switched'' field region experience a shift in the kinetic energy. This results in a change of the flight time to the detector, as shown in the upper panel of Fig. \ref{fig:NIE_Plot_2016}. Similar effects happen during the closing of the reflector after injection of the ions. The effect on the mass uncertainty is negligible for isobars, because they are not yet spatially separated.

From the green region in the upper panel in Fig. \ref{fig:NIE_Plot_2016}, the standard deviation is calculated and compared with the uncertainties of the individual measurement points in the plot. Based on this, $\Delta t_{NIE}$ is estimated. If the ions are outside of this region, they are discarded form the analysis. The uncertainty contribution is calculated by 

\begin{equation} \label{eq:NIE}
	\sigma_\mathrm{NIE} = \frac{2\Delta t_\mathrm{NIE}}{t_\mathrm{total\_IOI}} \cdot \sqrt{\frac{\left( \frac{m}{q}\right)_\mathrm{IOI}}{ \left( \frac{m}{q}\right)_\mathrm{REF}}} \cdot \left( \frac{m}{q}\right)_\mathrm{IOI} \; ,
\end{equation}
where $\sigma_\mathrm{NIE}$ is the uncertainty contribution due to non-ideal ejection, $ \left( \frac{m}{q}\right)_\mathrm{REF}$ is the mass-to-charge ratio of the ion used to obtain the $\Delta t_\mathrm{NIE}$ data, $t_\mathrm{total\_IOI}$ is the total TOF for the IOI and $\left( \frac{m}{q}\right)_\mathrm{IOI}$ is the mass-to-charge ratio of the IOI. 

The values of $\Delta t_\mathrm{NIE}$ were determined to be 0.1~ns and 0.5~ns for the data obtained in Experiments I \& II and Experiments III \& IV, respectively. The ions that travel towards the detector very close to the output reflector while it is being opened, experience an electrical ringing from the reflector electrodes. This is reflected in an oscillation in the TOF deviation for the longer opening delays, as shown in the grey area of the lower panel from Fig. \ref{fig:NIE_Plot_2016}. When the ions experience these fields, $\Delta t_\mathrm{NIE}$ will be increased to the peak value of the oscillation, which is about 2~ns for the example shown in Fig. \ref{fig:NIE_Plot_2016}.

\subsubsection{Uncertainty of resolved overlapping peaks}

The uncertainty from the correction for overlapping peaks is estimated to be 25~\% of the difference between the mass-to-charge ratio obtained with and without the correction, as described in section \ref{overlapping_peaks}.

\subsubsection{Unresolved-peaks uncertainty} \label{subsub:Unresolved_Peaks_Error}

In case of possible close-lying peaks of Class D, an additional uncertainty is added to the mass-to-charge value. When the expected peaks have a known distance but their abundance ratio is unknown, e.g. an unresolved isomer, isobar or an expected contamination, the estimated additional uncertainty is $\frac{\sqrt{3}}{6}$ times the expected mass-to-charge difference of the unresolved peaks \cite{Wang2017}. For the mass-to-charge ratios, a single-peak fit is performed and half of the known distance is added and subtracted to obtain the mass-to-charge values of the two unresolved peaks \cite{Wang2017}. 

When an unknown unresolved peak is probable, its effect is estimated with simulated data containing two peaks: one representing the IOI and the other a possible contamination. The abundance ratio between the IOI and the contamination is estimated by the highest unknown peak appearing with the same mass-to-charge number as the IOI. Then, the contaminant peak is moved over the IOI, calculating the mass-to-charge shift due to the influence of the contaminant. For each step a KS-test assuming a single peak is then performed to estimate whether such a contamination could be detected. For the uncertainty contribution, the maximum of mass-to-charge shift multiplied with the probability value of the KS-test is used, thereby taking into account the probability to identify the unknown peak.

\subsubsection{Phase-space uncertainty}\label{phase_space_un}
The phase space from the injection trap will be different for nuclei that have a different mass-to-charge ratio. These effects have been studied by computer experiments (SIMION, \cite{SIMION}). Also the effects of non-prefect voltages have been studied: (i) finite rise and fall times (~150~ns) of the push-pull voltages in the RF trap, (ii) RF fields in the trap during ejection, (iii) residual RF signals (~1V$_\mathrm{0p}$, capacitively coupled from the RF electrodes) at the push-pull apertures and (iv) imbalanced RF voltages (1\%) on the RF trap. For isobars and neighboring mass-to-charge ratios this is negligible. Even for relative mass-to-charge ratio differences as large as $\pm 25 \%$ the relative uncertainty does not exceed $2 \cdot  10^{-8}$ for (i,ii) and $2 \cdot  10^{-7}$ for (iii,iv). Since the RF is switched of shortly before ion ejection from the trap the effects of (ii-iv) are reduced by an order of magnitude.

\subsubsection{Uncertainty due to dead-time of data-acquisition system}
The TDC used for the data-acquisition, Ortec-9353, has a non-extending dead-time (1~ns) and the detector has an extending dead-time (0.5~ns). This means that after the detection of an event, there will be a window of time (1.5~ns in the case of Ortec-9353) where no event will be recorded. This effect results in an attenuation of the amplitude of the central part of the peak compared to the tails, thus altering the peak shape and increasing $\sigma$. A dead-time correction \cite{Greiner2017} is implemented in MAc to correct this uncertainty. The relative mass-to-charge uncertainty contribution due to the dead-time effects is less than $1 \cdot 10^{-8}$. This holds even for the condition of 0.5 ions detected on average per dead time. No uncertainty contribution has to be added in the data presented here.

\subsubsection{Space-charge uncertainty}
This contribution takes into account the interaction between isobaric ions while they travel close together in the analyzer. The magnitude of this effect in the relative mass-to-charge uncertainty is measured and described in \cite{Dickel2015}, and amounts to about $1\cdot10^{-8}$ per detected isobaric ion per MR-TOF-MS cycle. It can be neglected for all the data presented in this work.

\subsection{Final mass value from individual measurements} \label{sect:Data_Eval}
The mass-to-charge ratio and uncertainty obtained from the described data-analysis procedure can be converted to the mass value and uncertainty by Eq. \ref{eq:Multicharged_Mass} and Eq. \ref{eq:Multicharged_Sigma}, respectively:

\begin{equation} \label{eq:Multicharged_Mass}
m_\mathrm{IOI} = \left( \frac{m}{q}\right )_\mathrm{IOI} \cdot q_\mathrm{IOI} + m_\mathrm{e} \cdot q_\mathrm{IOI} 
\end{equation}

\begin{equation} \label{eq:Multicharged_Sigma}
\sigma_\mathrm{IOI} = \sigma_\mathrm{OI(m/q)} \cdot q_\mathrm{IOI}
\end{equation}
where $m_\mathrm{IOI}$ and $\sigma_\mathrm{IOI}$ are the final atomic mass and its uncertainty, $\left( \frac{m}{q}\right )_\mathrm{IOI}$ and $\sigma_\mathrm{IOI(m/q)}$ are the mass-to-charge ratio and its uncertainty obtained in the data-analysis procedure, $m_\mathrm{e}$ is the electron mass and $q_\mathrm{IOI}$ is the charge state of the measured IOI. Since measurements are performed on singly- or doubly-charged ions, the electron binding energies in the neutral atom can be neglected.

Combining results for the mass and abundance value of the same IOI, there are two cases: a) $N$ different measurements; b)  $N$ different analysis of the same data set. To determine the mass value a weighted mean is calculated according in Eq. \ref{eq:Mass_Combination}. It is assumed that all uncertainties are independent.

\begin{equation} \label{eq:Mass_Combination}
<m> = \frac{\sum_{i}^{N}\frac{1}{\sigma_\mathrm{i}^2}\cdot m_\mathrm{i}}{\sum_{i}^{N}\frac{1}{\sigma_\mathrm{i}^2}} \; ,
\end{equation}
where $m_\mathrm{i}$ and $\sigma_\mathrm{i}$ are the individual values and their uncertainties,  respectively. For case a) the uncertainty is divided in an independent component, calculated via the variance of the weighted mean, and a dependent one, calculated via the weighted mean.  The uncertainty components are added quadratically. In case of b) the weighted mean of the uncertainties is used.

\section{Results} 
The data-analysis procedure and the MR-TOF-MS system have been used for the four experiments described in Section 2. The measured results presented here, cover the following topics: (i) accuracy of the procedure with isotopes of well-known masses, (ii) first-time direct mass measurements, and (iii) first-time measurements of isomer-to-ground state ratios.

\subsection{Mass accuracy}
\begin{figure}[htb]
\centering 
	\includegraphics[width=0.9\linewidth]{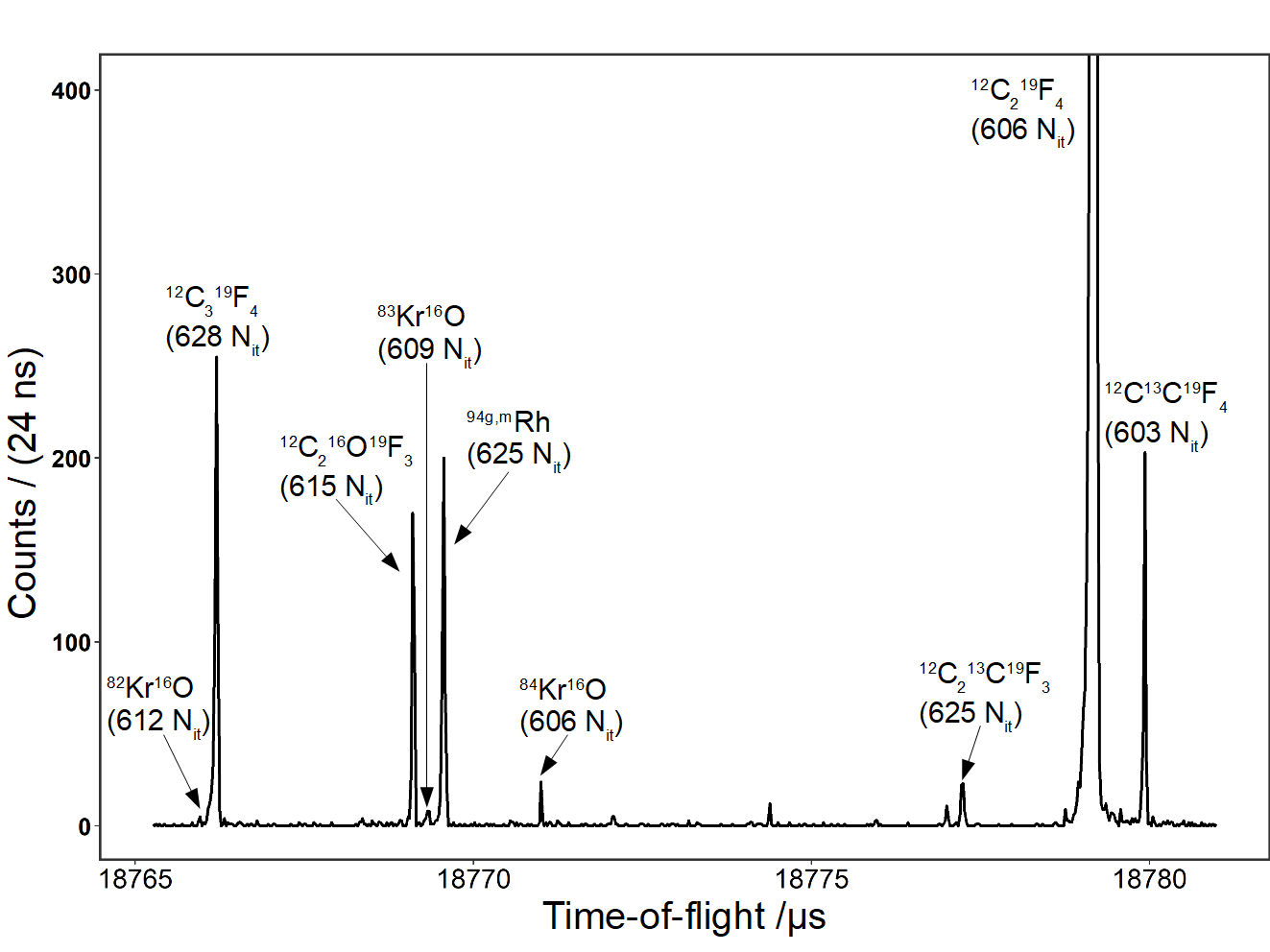}
\caption{Time-of-flight spectrum of different atomic and polyatomic ions with TOF of about 18.8~ms. Depending on their mass-to-charge ratio, the ions undergo 603 to 628 isochronous turns ($N_{it}$).}
\label{fig:MT-spectrum2016_tof}
\end{figure}

The broadband characteristic of the MR-TOF-MS enables simultaneous measurement of exotic nuclei with different mass ($A$) and element numbers ($Z$). This allows very efficient measurements. As a consequence the broadband measurements with the highest resolving powers yield complex mass-to-charge spectra. However, the data-analysis procedure developed here is capable of analyzing these spectra. An example is given in Fig. \ref{fig:MT-spectrum2016_tof}. 

\begin{figure}[htb]
	\centering 
	\includegraphics[width=0.9\linewidth]{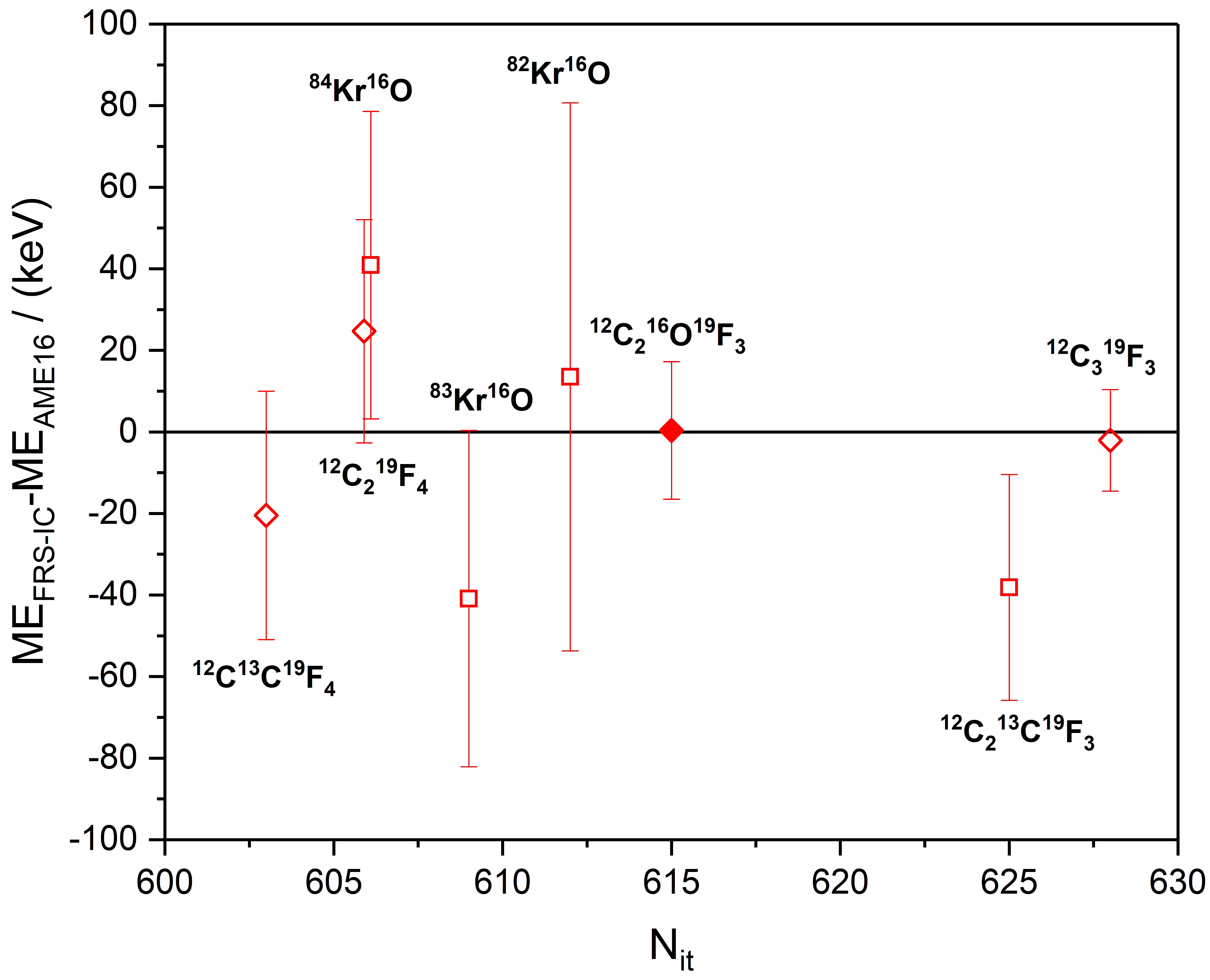}
	\caption{Deviations of measured mass excess (ME) values from literature values, obtained for ions undergoing different numbers of turns ($N_\mathrm{it}$). The corresponding TOF spectrum in shown in Fig.~\ref{fig:MT-spectrum2016_tof}. The ions $^{12}$C$^{13}$C$^{19}$F$_{4}^{1+}$, $^{12}$C$_{2}^{19}$F$_{4}^{1+}$, $^{12}$C$_{2}^{16}$O$^{19}$F$_{4}^{1+}$ and $^{12}$C$_{3}^{19}$F$_{3}^{1+}$ (open diamond symbol) were used for the determination of the calibration parameters $t_0$ and $c$. TRC was performed using the ion $^{12}$C$_{2}^{13}$C$^{19}$F$_{3}^{1+}$. The ion $^{12}$C$_{2}^{16}$O$^{19}$F$_{3}^{1+}$ (filled diamond symbols) was used for the calculation of the final mass-to-charge value.}
	\label{fig:MT-spectrum2016}
\end{figure}

The species shown have a maximum difference of 25 isochronous turns and cover about 10~u/e. The measured Mass Excesses (ME) values are compared with the literature values in dependence of the number of turns in Fig. \ref{fig:MT-spectrum2016}.

The experimental results agree with the literature values with typical uncertainties of 20~keV. Note, that this uncertainty is merely the result for the example presented here and higher accuracies can be achieved for spectra for different turn numbers.
 
Nuclei with masses known to high accuracy were studied to determine the mass accuracy of the system and the data-analysis procedure. These include low-lying isomers with low statistics, thus representing the most challenging cases for the data-analysis procedure.

\begin{figure}[htp]
\centering
\includegraphics[width=0.9\linewidth]{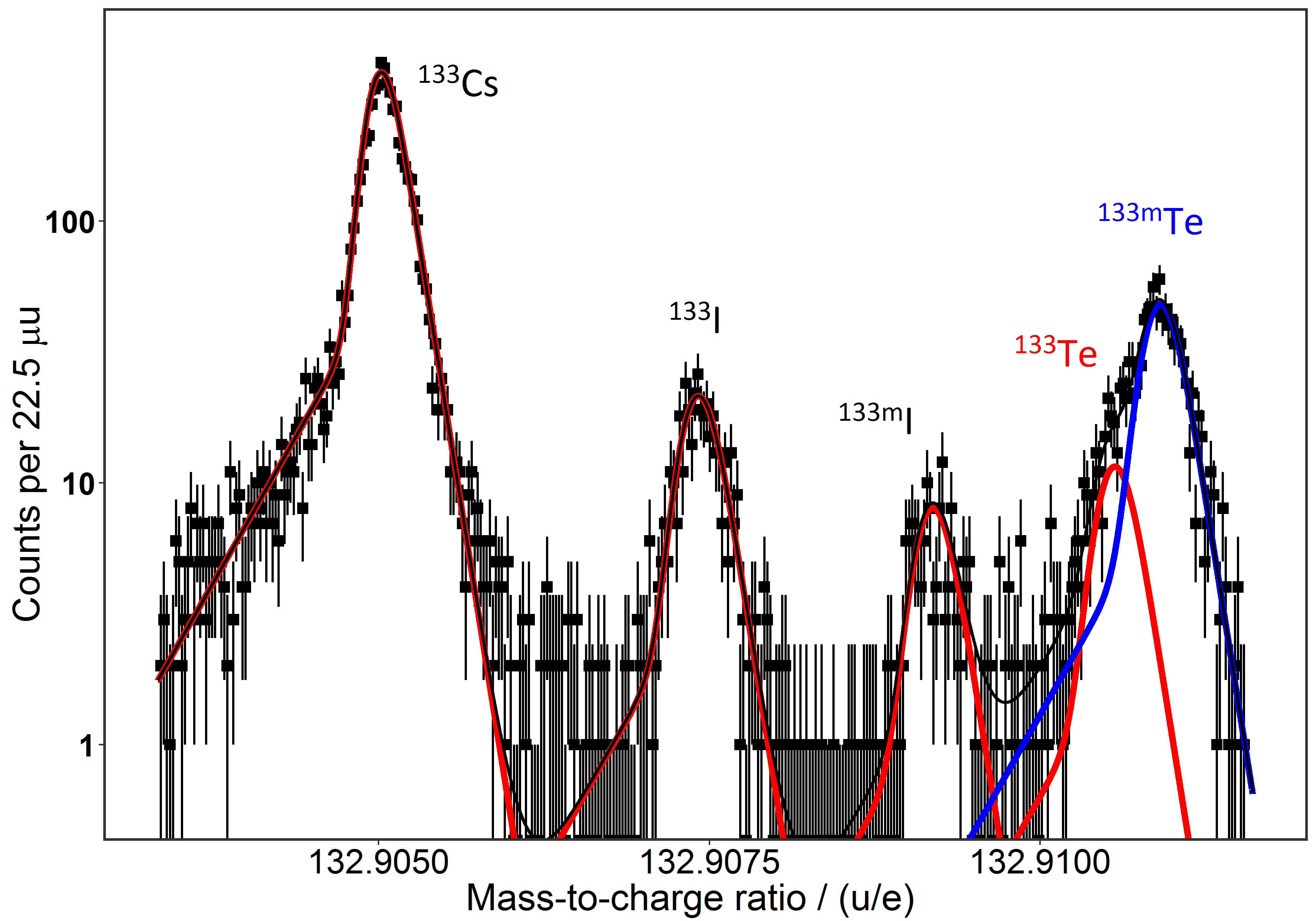}
\caption[$^{133}$I, $^{133m}$I and $^{133}$Te, $^{133m}$Te Full Mass Spectrum]{Mass-to-charge spectrum of the nuclides $^{133}$I, $^{133m}$I (excitation energy of about 1.6~MeV) and $^{133}$Te, $^{133m}$Te (excitation energy of about 300~keV) obtained with the MR-TOF-MS in Experiment I. The binned data (black squares), the fits with the Hyper-EMG(1,1) function (red and blue curves), and the sum of all fits (black curve) are shown. The spectrum was calibrated using $^{133}$Cs ions. The mass resolving power (FWHM) amounts to 410,000. Note that the mass-to-charge spectrum is shown with a logarithmic aboundance scale. }
\label{fig:FRSIC14_133Te_Full}
\end{figure}

In Fig. \ref{fig:FRSIC14_133Te_Full}, a spectrum taken with the MR-TOF-MS during Experiment II and analyzed with the procedure described above is shown. In this spectrum, the calibrant and two nuclides with their long-lived isomeric states are seen, whose masses and excitation energies are well-known from literature.

The acquisition time for this measurement was about 2~hours. The excitation energies of the isomers are 1634~keV and 334~keV for iodine and tellurium, respectively. In the case of iodine, the ground and isomeric states can be clearly resolved. For tellurium, the ground and isomeric states are overlapping, and the isomer is visible only as a shoulder on the left side of the isomeric peak. Our data-analysis procedure enabled extraction of the masses and abundance ratios also in the most challenging case.

\begin{figure*}[htp]
\centering
\includegraphics[width=1.\linewidth]{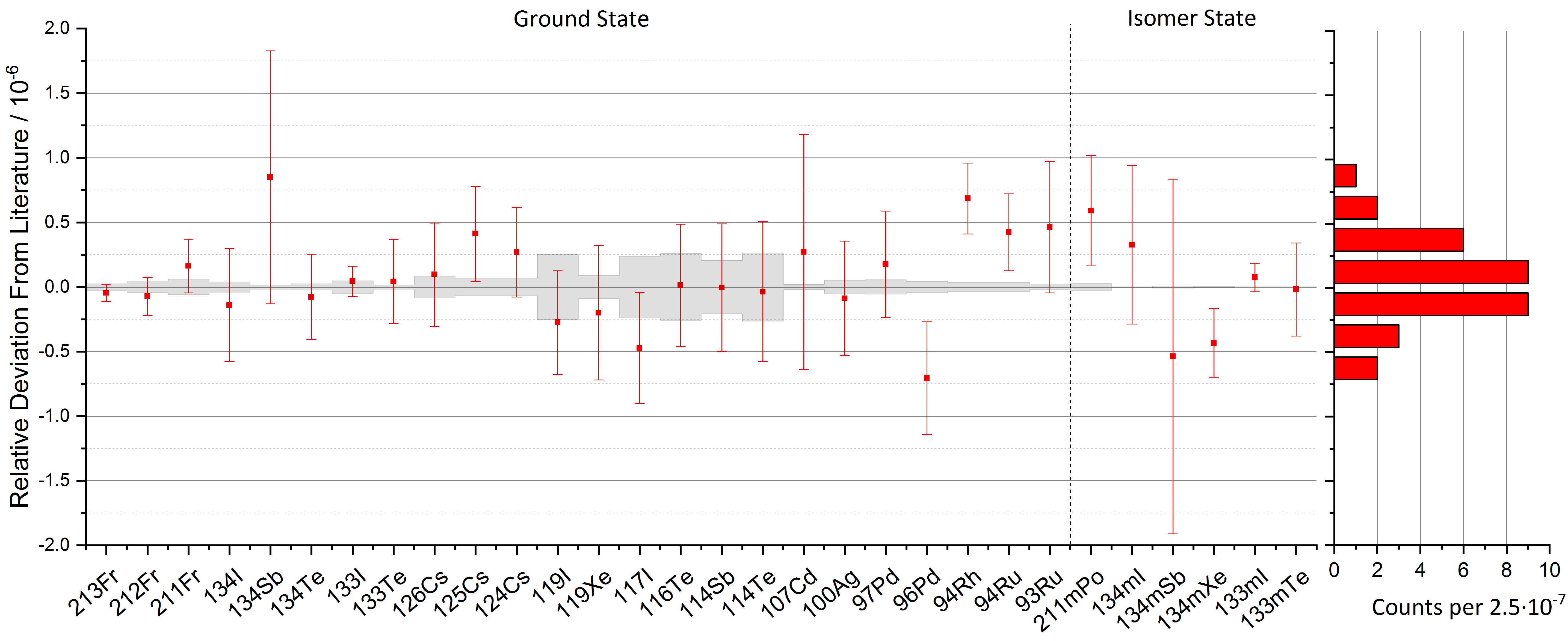}
\caption[Compilation Plot]{Relative deviation of measured ground state masses $(m_\mathrm{FRS-IC}-m_\mathrm{AME16})/m_\mathrm{FRS-IC}$ (table~\ref{tab:FRSIC_Masses}) and isomer excitation energies $(E_\mathrm{ex,FRS-IC}-E_\mathrm{ex,AME16})/m_\mathrm{FRS-IC}$ (table~\ref{tab:FRSIC_Isomers}) from the literature values given in AME2016 \cite{Wang2017}. All relative deviations are calculated with respect to the ground state masses. The grey band around the horizontal axis represents the literature uncertainty. In the right-hand panel, a histogram of the relative deviations with a bin size of $2.5\cdot 10^{-7}$ is shown. The weighted mean of all relative deviations is $(4.5 \pm 5.3)\cdot 10^{-8}$ and the standard deviation is $3.5 \cdot 10^{-7}$. The Birge ratio \cite{Birge1932} of the values is 0.891, showing that the different sources of uncertainty presented here describe adequately the total uncertainty, and that no unknown systematic uncertainty needs to be added. For 6 out of 31 nuclei a deviation larger one standard deviation occurs, this is less than expected form a normal distributed uncertainties.}
\label{fig:FRSIC_Compilation}
\end{figure*}

The next case shows overlapping peaks with very low statistics, placing an even harder challenge for the data-analysis procedure.
\begin{figure}[htb]
\centering
\includegraphics[width=0.9\linewidth]{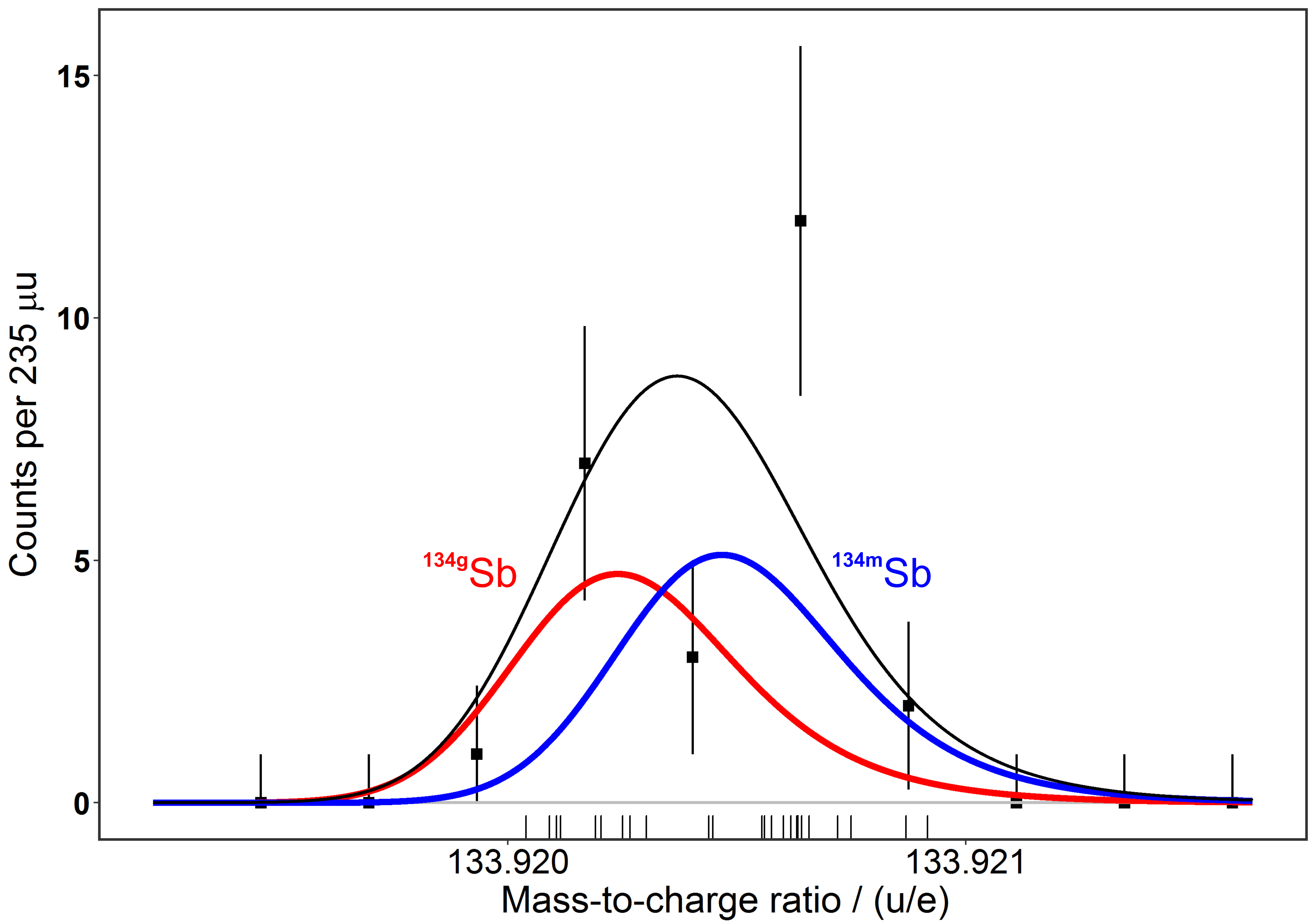}
\caption[$^{134}$Sb, $^{134m}$Sb Hyper-EMG Fitted Mass Spectrum]{Mass-to-charge spectrum of $^{134}$Sb and $^{134m}$Sb ions. There are 25 counts in the spectrum. The unbinned data events are depicted above the mass-to-charge ratio axis. The curves represent a fit with Hyper-EMG(0,1) functions to the unbinned data. The curves correspond to $^{134}$Sb (red), $^{134m}$Sb (blue) and their sum (black).}
\label{fig:Peakform_134Sb_134mSb_e01_MLE_lin}
\end{figure}
Due to a detection rates of only 12 events per hour, only 25 counts were available for both the ground and isomeric state, expected at an excitation energy of 279~keV. A single-peak fit to the data of $^{134}$Sb and $^{134m}$Sb, with peak-shape parameters obtained from the calibrant peak (Hyper-EMG(0,1)), resulted in a mass value in between the isomer and ground state with a p-value of the KS test of 0.29. When performing a double-peak fit with two Hyper-EMG(0,1) functions, a higher p-value of 0.42 and mass values of the ground state and the isomeric state consistent with literature mass were obtained, with relative uncertainties below 10$^{-6}$ (see Fig.\ref{fig:FRSIC_Compilation} and table \ref{tab:FRSIC_Masses}). 

The result of the double-peak fit to the data is shown in Fig. \ref{fig:Peakform_134Sb_134mSb_e01_MLE_lin}. In addition to the ground state mass and isomer excitation energy, the abundance ratio has been determined to be $1.08 \pm 0.73$. The uncertainty is dominated by the fact that the peaks are overlapping. If they were resolved, an uncertainty of about 0.4 would be expected just from statistics. This level of uncertainty (mass, excitation energy and isomer-to-ground state ratio) with such a low number of events is possible only due to the high resolving power of the MR-TOF-MS and the data-analysis procedure described here.

\begin{table*}[htp]
\centering
    \begin{tabular}{ cccccccc } \hline\hline
    Nuclide & Half-life & Exp. & Reference ion & ME$_\mathrm{FRS-IC}$ & ME$_\mathrm{AME16}$ & ME$_\mathrm{FRS-IC}$ - ME$_\mathrm{AME16}$ & Number \\
              &        &               &      &   / keV &  / keV & / keV  & of events \\ \hline
    	$^{213}$Fr & 34.14 $\pm$ 0.06$\>$s& I & $^{211}$Pb & -3561.9 $\pm$ 12.3 & -3553 $\pm$ 5 & -8.9 $\pm$ 13.3 & 6022\\
    	$^{212}$Fr & 20.0 $\pm$ 0.6$\>$min& I & $^{211}$Pb & -3530 $\pm$ 28 & -3516 $\pm$ 9 & -14 $\pm$ 29 & 445 \\
    	$^{211}$Fr & 3.10 $\pm$ 0.02$\>$min& I & $^{211}$Pb & -4108 $\pm$ 40 & -4140 $\pm$ 12 & 32 $\pm$ 41 & 126 \\
        $^{134}$I & 52.5 $\pm$ 0.2$\>$min & II & $^{134}$Xe & -84062 $\pm$ 54 & -84043 $\pm$ 5 & -19 $\pm$ 54 & 172 \\
    	$^{134}$Sb & 780 $\pm$ 60$\>$ms & II & $^{134}$Xe & -73915 $\pm$ 122 & -74021 $\pm$ 2 & 106 $\pm$ 122 & 12    \\
     	$^{134}$Te & 41.8 $\pm$ 0.8$\>$min & II & $^{134}$Xe & -82543 $\pm$ 41 & -82534 $\pm$ 3 & -10 $\pm$ 41 &  61   \\
    	$^{133}$I & 20.83 $\pm$ 0.08$\>$h & II & $^{133}$Cs & -85852 $\pm$ 15  & -85858 $\pm$ 6 & 5 $\pm$ 15 &  566  \\
        $^{133}$Te & 12.5 $\pm$ 3$\>$min & II & $^{133}$Cs & -82932 $\pm$ 40 & -82937 $\pm$ 2 & 5 $\pm$ 40 & 423    \\ 
        $^{126}$Cs & 1.64 $\pm$ 0.02$\>$min & III & $^{126}$Xe & -84340 $\pm$ 46 & -84351 $\pm$ 10 & 11 $\pm$ 47 & 22    \\
        $^{125}$Cs & 46.7 $\pm$ 0.1$\>$min & III & $^{126}$Xe & -84040 $\pm$ 42 & -84088 $\pm$ 8 & 48 $\pm$ 43 & 609    \\
        $^{124}$Cs & 30.9$\pm$ 0.4$\>$s & III & $^{126}$Xe & -81700 $\pm$ 39 & -81269 $\pm$ 8 & 31 $\pm$ 40 & 23    \\
       	$^{119}$I & 19.1 $\pm$ 0.4$\>$min & III & $^{12}$C$_{2}$ $^{19}$F$_{5}$ (A=119) & -83796 $\pm$ 34 & -83766 $\pm$ 28 & -30 $\pm$ 44 & 90    \\
       	$^{119}$Xe & 5.8 $\pm$ 0.3$\>$min & III & $^{12}$C$_{2}$ $^{19}$F$_{5}$ (A=119) & -78816 $\pm$ 57 & -78794 $\pm$ 10 & -22 $\pm$ 58 & 31    \\
       	$^{117}$I & 2.22 $\pm$ 0.04$\>$min & III & $^{12}$C$_{2}$ $^{19}$F$_{5}$ (A=119) & -80488 $\pm$ 47 & -80436 $\pm$ 26 & -51 $\pm$ 54 & 1022    \\
       	$^{116}$Te & 2.49 $\pm$ 0.04$\>$h & III & $^{12}$C$_{2}$ $^{19}$F$_{5}$ (A=119) & -85268 $\pm$ 51 & -85269 $\pm$ 28 & 1 $\pm$ 58 & 1183    \\
       	$^{114}$Sb & 3.49 $\pm$ 0.03$\>$min & III & $^{12}$C$_{2}$ $^{19}$F$_{5}$ (A=119) & -84497 $\pm$ 47 & -84497 $\pm$ 22 & -1 $\pm$ 52 & 347    \\
        $^{114}$Te & 15.2 $\pm$ 0.7$\>$min & III & $^{12}$C$_{2}$ $^{19}$F$_{5}$ (A=119) & -81893 $\pm$ 50  & -81889 $\pm$ 28 & -4 $\pm$ 58 & 269    \\
       	$^{107}$Cd & 6.50 $\pm$ 0.02$\>$h & IV & $^{84}$Kr $^{14}$N$_{2}$ (A=112) & -86963 $\pm$ 90 & -86990 $\pm$ 2 & 27 $\pm$ 90 & 47    \\
        $^{100}$Ag & 2.01 $\pm$ 0.09$\>$min & IV & $^{12}$C$_{2}$ $^{19}$F$_{4}$ (A=100) & -78146 $\pm$ 41 & -78138 $\pm$ 5 & -8 $\pm$ 41 & 36$^{*}$     \\
        $^{97}$Pd & 3.10 $\pm$ 0.09$\>$min& IV &  $^{12}$C$_{2}$ $^{16}$O $^{19}$F$_{3}$ (A=97) & -77790 $\pm$ 37 & -77806 $\pm$ 5 & 16 $\pm$ 37 & 35 \\
        $^{96}$Pd & 122 $\pm$ 2$\>$s& IV & $^{12}$C$_{2}$ $^{19}$F$_{4}$ (A=100) & -76246 $\pm$ 38 & -76183 $\pm$ 4 & -63 $\pm$ 39 & 224 \\
        $^{94}$Rh & 70.6 $\pm$ 0.6$\>$s$^{**}$& IV & $^{12}$C$_{2}$ $^{16}$O $^{19}$F$_{3}$ (A=97) & -72848 $\pm$ 24 & -72908 $\pm$ 3 & 60 $\pm$ 24 & 338 \\
        $^{94}$Ru & 51.8 $\pm$ 0.6$\>$min& IV & $^{12}$C$_{3}$ $^{13}$C $^{19}$F$_{3}$ (A=94)& -82547 $\pm$ 26 & -82584 $\pm$ 3 & 37 $\pm$ 26 & 88 \\
        $^{93}$Ru & 59.7 $\pm$ 0.6$\>$s& IV &  $^{12}$C$_{2}$ $^{16}$O $^{19}$F$_{3}$ (A=97) & -77177 $\pm$ 44 & -77217 $\pm$ 2 & 40 $\pm$ 44 & 20 \\ \hline\hline
        
    \end{tabular}
\caption[Mass Measurements FRS-IC 2014 and 2016 Experiments]{Results of direct mass measurements performed in the FRS-IC in Experiments I - IV. The shown uncertainties are the total experimental uncertainty. In the Experiment III the uncertainties are often dominated by higher background, see section \ref{sc_exp}. Literature values are from \cite{Audi2016,Wang2017}. $^{*}$ Total number of events for the unresolved ground and isomeric states. $^{**}$ Assignment to ground state and isomer is uncertain. }
\label{tab:FRSIC_Masses}
\end{table*}
\renewcommand{\arraystretch}{1}

All the masses and excitation energies measured and listed in table \ref{tab:FRSIC_Masses} and table \ref{tab:FRSIC_Isomers} have been directly measured previously by techniques such as Penning trap mass spectrometry (TOF-ICR) or Schottky Mass Spectrometry in storage rings. Therefore, they can be used as references to test possible systematic shifts and unknown systematic uncertainties of the spectrometer or data-analysis procedure presented in this work. 

A histogram of the relative deviations between the masses and excitation energies obtained with the MR-TOF-MS of the FRS-IC and the literature values \cite{Wang2017} is shown in the right side of Fig. \ref{fig:FRSIC_Compilation}, with a weighted mean of ($4.5 \pm 5.3)\cdot10^{-8}$ and a standard deviation of $3.5\cdot10^{-7}$. This clearly demonstrates that the system and the data-analysis procedure provide highly accurate mass values. Note, about half of the masses have been measured with less than 100 events, and about a quarter of the peaks are not baseline-resolved, i.e.\ they belong to Class B, C or D.

For $^{100}$Ag it is known to have an low-lying isomeric states with excitation energies of 15.5~keV, which were not resolved in the measurement. Their corresponding half-lives are similar on the time scale of the measurement. Therefore it is assumed to be measured as a mixture of both states. The ground state masses are evaluated as described in section \ref{subsub:Unresolved_Peaks_Error}.

\subsection{Isomer-to-ground state ratios}
Isomer-to-ground state ratio shed light on fragmentation in peripheral heavy-ion collisions at relativistic energies, guiding the efficient use of in-flight fragmentation at future radioactive beam facilities \cite{Bowry2013}. When nuclides are produced by fission, isomer-to-ground state ratio provide insights regarding the origin of the angular momentum of the fission fragments, which in turn may reveal properties of the dynamical evolution of the fissioning nucleus from the saddle point until its descent to scission \cite{rakopoulos2018}. 

The ratios of isomer-to-ground state measured with the MR-TOF-MS are given in table \ref{tab:FRSIC_Isomers}. The isomer-to-ground state ratio of $^{134}$Sb ions has now been studied in $^{235}$U thermal neutron-induced fission \cite{kerek1972}, $^{232}$Th 25-MeV proton induced fission \cite{kankainen2013}, $^{252}$Cf spontaneous fission \cite{siegl2018} and now with $^{238}$U abrasion-fission.
\renewcommand{\arraystretch}{1.15}
\begin{table*}[htp]
\centering
    \begin{tabular}{cccccccc} \hline\hline
    Isomer & $J^{\pi}$ & Half-life & Exp. & E$_\mathrm{ex,FRS-IC}$ & E$_\mathrm{ex,LIT}$ & E$_\mathrm{ex,FRS-IC}$ - E$_\mathrm{ex,LIT}$ & Isomer-to-ground \\ 
                     &           &   &  & / keV & / keV & / keV & state ratio \\ \hline
	$^{211m}$Po	& $25/2^+$	& 25.2 $\pm$ 0.6$\>$s& I & 1578 $\pm$ 84		& 1462 $\pm$ 5	   & 116 $\pm$ 84 & 3.30 $\pm$ 1.42	\\
    $^{134m}$I 	& $8^-$	& 3.52 $\pm$ 0.04$\>$min & II & 357 $\pm$ 76				& 316.49 $\pm$ 0.22	& 41 $\pm$ 76 & 1.26 $\pm$ 0.26  	\\
    $^{134m}$Sb	& $7^-$	& 10.07 $\pm$ 0.05$\>$s & II & 212 $\pm$ 171		& 279 $\pm$ 1 	    & -67 $\pm$ 171 & 1.08 $\pm$ 0.73 	\\
    $^{134m}$Xe & $7^-$	& 290 $\pm$ 17$\>$ms & II & 1911 $\pm$ 34			& 1965.5 $\pm$ 0.5	& -55 $\pm$ 34 & 0.035 $\pm$ 0.006	\\
    $^{133m}$I 	& $19/2^-$	& 9 $\pm$ 2$\>$s & II & 1643 $\pm$ 14		& 1634.148 $\pm$ 0.010	& 9 $\pm$ 14 & 0.35 $\pm$ 0.02 	\\
    $^{133m}$Te	& $19/2^-$ 	& 55.4$\pm$ 0.4$\>$min & II & 332 $\pm$ 45			& 334.26 $\pm$ 0.04	& -2 $\pm$ 45 & 4.12 $\pm$ 0.26 	\\ \hline\hline
    \end{tabular}
\caption[Isomers]{Excitation energies and isomer-to-ground state ratios measured with the MR-TOF-MS. The isomer-to-ground state ratios are not corrected for decay losses, because the measurement time is short compared to the half-live of the studied states. Literature values are from \cite{Audi2016,Jain2015}.}
\label{tab:FRSIC_Isomers}
\end{table*}
\renewcommand{\arraystretch}{1}

\subsection{Masses of short-lived nuclides directly measured for the first time}
Above the doubly-magic nucleus $^{208}$Pb a region of very short-lived nuclei opens up in the nuclear chart with half-lives down to nanoseconds. The masses of these nuclei were measured so far only via their $Q$-values. They relate the mother and daughter nuclei masses via $\alpha$-decay. This is only unambiguous if the initial and final states are known. Therefore, direct mass measurements are desirable for the $\alpha$-emitters, such as the recent mass measurements performed at RIKEN \cite{Rosenbusch2018}. In the following, we present first-time direct mass measurements of seven $\alpha$-emitters in this region of interest. The measured results are listed in table \ref{tab:FRSIC_Direct_Masses}.

\begin{table*}[t]
\centering
    \begin{tabular}{cccccccc} \hline\hline
    Nuclide & Half-life & Exp. & Reference ion & ME$_\mathrm{FRS-IC}$ & ME$_\mathrm{AME16}$ & ME$_\mathrm{FRS-IC}$ - ME$_\mathrm{AME16}$ & Number \\
     &  & & & / keV & / keV & / keV & of events \\ \hline
		$^{220}$Ra & 17.9 $\pm$ 1.4$\>$ms & I & $^{34}$S $^{19}$F$_{4}$ (A=110) & 10609 $\pm$ 320 & 10270 $\pm$ 8 & 339 $\pm$ 320 & 11 \\
        $^{218}$Rn & 33.75 $\pm$ 0.15$\>$ms& I & $^{219}$Rn & -5089 $\pm$ 54 & -5217.3 $\pm$ 2.3 & -128 $\pm$ 55 & 162 \\ 
        $^{217}$At & 32.62 $\pm$ 0.24$\>$ms& I & $^{219}$Rn & 4433 $\pm$ 135 & 4395 $\pm$ 5 & 38 $\pm$ 135 & 19 \\
        $^{213}$Rn & 19.5 $\pm$ 0.1$\>$ms& I & $^{211}$Pb & -5737 $\pm$ 63$^{*}$ & -5696 $\pm$ 3 & -41 $\pm$ 63$^{*}$ & 165 + 29$^{*}$ \\
        $^{212}$At & 314 $\pm$ 2$\>$ms& I & $^{211}$Pb & -8601 $\pm$ 86 & -8628.2 $\pm$ 2.4 & -27 $\pm$ 86 & 1496$^{**}$ \\
        $^{212}$Rn & 23.9 $\pm$ 1.2$\>$min& I & $^{211}$Pb & -8609 $\pm$ 30 & -8660 $\pm$ 3 & 51 $\pm$ 30 & 514 \\
        $^{211}$Po & 516 $\pm$ 3$\>$ms& I & $^{211}$Pb & -12593 $\pm$ 137$^{*}$ & -12432.6 $\pm$ 1.3 & -160 $\pm$ 137$^{*}$ & 78 + 411$^{*}$ \\ \hline\hline
 
    \end{tabular}
\caption[First Direct Mass Measurements FRS-IC 2014 and 2016 Experiments]{Results of first-time direct mass measurements performed at the FRS-IC. The uncertainties shown correspond to the total experimental uncertainty. Literature values are from \cite{Audi2016,Wang2017}. $^{*}$ Weighted mean value including the results of a previous experiment at the FRS-IC \cite{Ebert2016,Jesch2016}. $^{**}$ Total number of events for the unresolved ground and isomeric states.}
\label{tab:FRSIC_Direct_Masses}
\end{table*}
\renewcommand{\arraystretch}{1}

\subsubsection{$^{212,\ 213,\ 218}$Rn isotopes}
The $^{212}$Rn nucleus has a closed shell with 126 neutrons and a half-life of 23.9~min, the neighboring isotope $^{213}$Rn has a much shorter half-life of 19.5~ms and the $^{218}$Rn nucleus has a half-life of 33.75(15)~ms. 

The present mass measurement was performed with 128 isochronous turns, corresponding to a TOF of about 5.8~ms and a mass resolving power of about 200,000. 
The masses of these three nuclides were determined in the past by $\alpha$-decay ($^{212}$Rn:\cite{Momyer1955,Golovkov1971}; $^{213}$Rn:\cite{Valli1967,Valli1970}; $^{218}$Rn:\cite{Bowman1982,Asaro1956})

\subsubsection{$^{211}$Po isotope}
The ground and isomeric states of $^{211}$Po nuclei have been studied by $\alpha$-spectroscopy (eg., \cite{Jentschke1954,Walen1962b,Bowman1982}) in the past. 
In our mass measurement, the $^{211}$Po ions traveled 192 isochronous turns in the analyzer of the MR-TOF-MS, resulting in a mass resolving power of 300,000. This is not enough to resolve the ground state of $^{211}$Pb and the isomeric state of $^{211}$Po, separated by 574~keV/c$^2$. The radioactive ion source installed in the CSC has produced the $^{211}$Pb ions. They were used for calibration at the beginning and end of the measurement, when no beam entered the CSC. During operation with beam, the $^{211}$Pb ion source was blocked by an electric field (220~V, DC), but some $^{211}$Pb background ions remained. This was taken into account in the data analysis by including a third peak for $^{211}$Pb. The parameters for this peak was fixed. The uncertainty of the parameters of the third peak, especially the rate of the $^{211}$Pb background ions, has been considered for the final mass uncertainty. For the $(25/2 )^{+}$-isomer of $^{211}$Po an excitation energy of 1578(84)~keV was measured which is in agreement with our previously direct measurement 1472 (120)~keV \cite{Dickel2015b}, the literature value is 1462(5)~keV \cite{Jain2015}.

\subsubsection{$^{220}$Ra isotope}
The measurement and data analysis of $^{220}$Ra ions was especially challenging, because: (i) $^{220}$Ra nuclides has a half-life of 17.9 ms, which is the shortest-lived isotope measured with an MR-TOF-MS up to now, (ii) the ion was measured as doubly-charged, (iii) only 11 counts were recorded. The MR-TOF-MS and the data-analysis procedures described in this paper have been developed to cope with these challenges.

\subsubsection{$^{212,\ 217}$At isotopes}

The $^{212}$At nucleus has a half-life of 314~ms for the $( 1 )^{-}$ ground state and 119~ms for the $( 9 )^{-}$ isomeric state with an excitation energy of 222.9(0.9) keV. This mass difference was not resolved in the measurement. Both states decay via $\alpha$-decay. In this measurement, the FRS was set to the very short-lived fragment $^{216}$Fr, which was stopped in the CSC and decayed into $^{212}$At via $\alpha$-decay. Both nuclei, $^{216}$Fr and $^{212}$At, have a $( 9)^{-}$ isomeric state. The $( 9)^{-}$~isomer of $^{216}$Fr decays in similar manner to its ground state $( 1)^{-}$ via $\alpha$-decay \cite{Kurcewicz2007}. The ground state and the $( 9)^{-}$~isomer have similar half-lives - 700~ns and 850~ns, respectively. Therefore both states decay in the CSC, the recoil energy is absorbed by the helium gas, and the ground and isomeric state of $^{212}$At are populated. 
Kurcewicz et al. \cite{Kurcewicz2007} reported an isomeric ratio for the $( 9)^{-}$ state of 0.28(1) for $^{212}$At and 0.31(2) for $^{216}$Fr. The latter nucleus was populated via $\alpha$-decay of $^{220}$Ac. 
In a later investigation \cite{Wojtasiewicz2009} the isomeric ratio for the $(9)^{-}$ state in $^{212}$At was again determined. The obtained value was 0.09(2) which differs strongly from the reported value of reference \cite{Kurcewicz2007}. 

$^{217}$At has a half-life of 32.3~ms. Its production cross section is 18 $\mu$barn, the smallest of all isotopes presented here. Therefore, the uncertainty in its mass measurement is dominated by statistics, as the spectrum included only 19 ions.
The mass of $^{217}$At nuclides was measured previously only by $\alpha$-spectroscopy \cite{Vorob1960,Bowman1982}. 

\section{Experimental results compared with theoretical predictions at the lead shell closures}
Accurate mass values and also their differences can provide basic information of the strong interaction in nuclei and are essential for the understanding of the synthesis of the elements in the universe \cite{Schatz2013}. Nuclear shell stabilization is the reason for the existence of the  heaviest experimentally known elements and particularly for the superheavy elements \cite{Oganessian2001,muenzenberg2015}. The doubly magic nucleus $^{208}$Pb, formed by bound 82 protons and 126 neutrons, represents the heaviest experimentally and theoretically well-known proton shell closure. The masses in this domain have been experimentally determined mainly via decay data and are now for the first time directly measured with the MR-TOF-MS. Therefore, this region of nuclides is an ideal testing ground for our new experimental method discussed in this work.      
The new mass values, measured with the MR-TOF-MS, have been applied, together with the experimentally known data, to investigate the accuracy of different basic theoretical predictions in the Pb region.  

\begin{figure}[htp]
\centering
\includegraphics[width=0.9\linewidth]{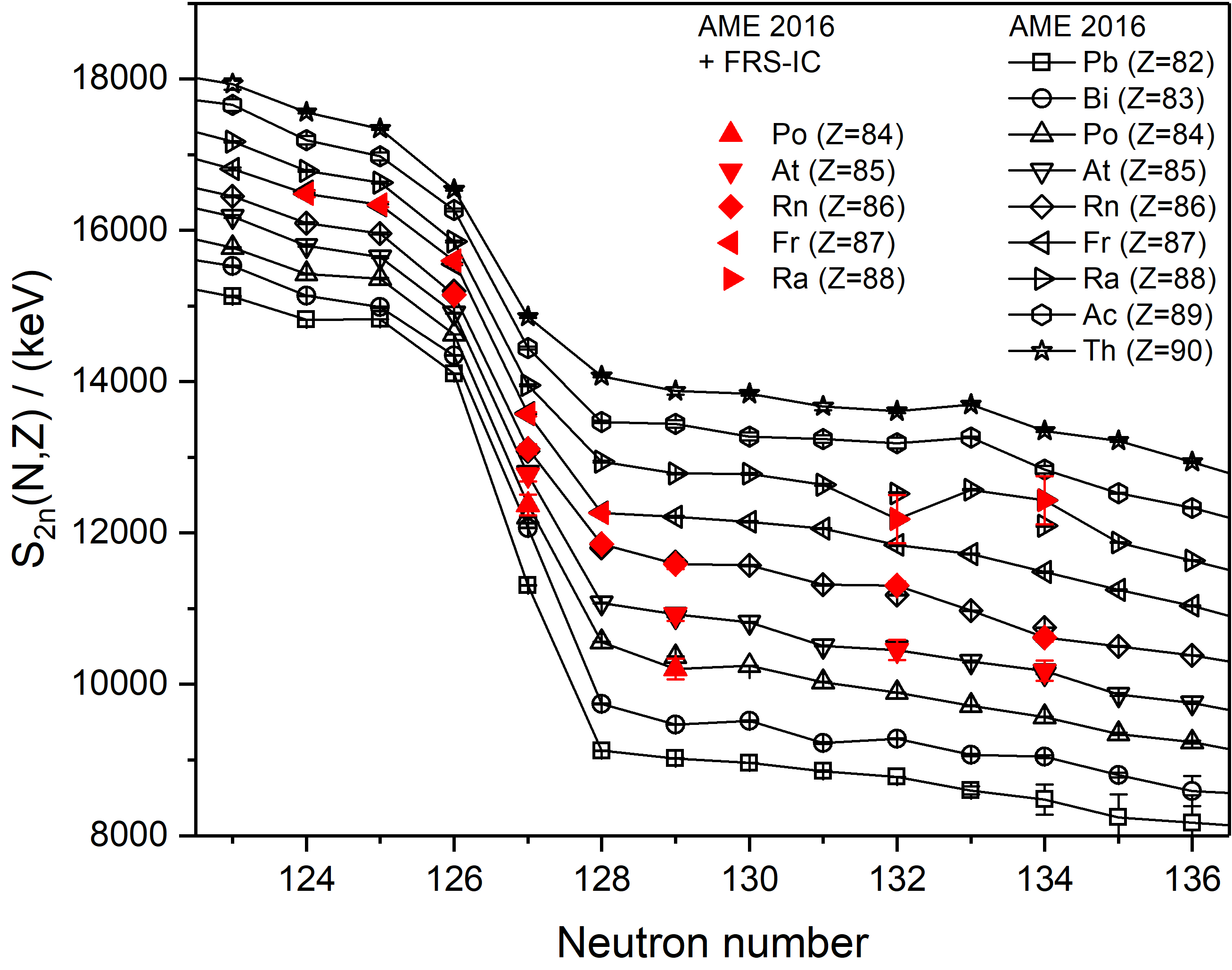}
\caption[S2N]{Two neutron separation energy $S_\mathrm{2n}$ versus neutron number including mass excess data from AME16 (open symbols) and the ones measured in this work (red filled symbols). The closed neutron shell at N=126 appears as a steeper slope.}
\label{fig:S2N}
\end{figure}

The two-neutron separation energy ($S_\mathrm{2n}$) is defined by the relation of the mass excess values (ME):
\begin{equation} \label{eq:S2N}
S_\mathrm{2n}\left(N,Z\right) = ME\left(N-2,Z\right) + 2~ME\left(n\right) - ME\left(N,Z\right) \; ,
\end{equation}
where $ME\left(n\right)$ is the mass excess of the neutron. The $S_\mathrm{2n}$ surface in the Pb mass region  is shown in Fig. \ref{fig:S2N}. 

As already presented in the previous chapters, the new MR-TOF-MS data are in excellent agreement with previously known data which have been determined via Q$_\alpha$ values and other quite different experimental methods. 
The experimental $S_\mathrm{2n}$ values show the expected steep drop at the N=126 shell  for all elements in the region. A closer look indicates that the slope becomes shallower with an increasing difference of the proton number from the closed shell Z=82. 

This evolution of the shell gap near the double magicity can be illustrated by the slope of the difference of the two-neutron separation energies for different elements.  

\begin{figure}[htp]
\centering
\includegraphics[width=0.9\linewidth]{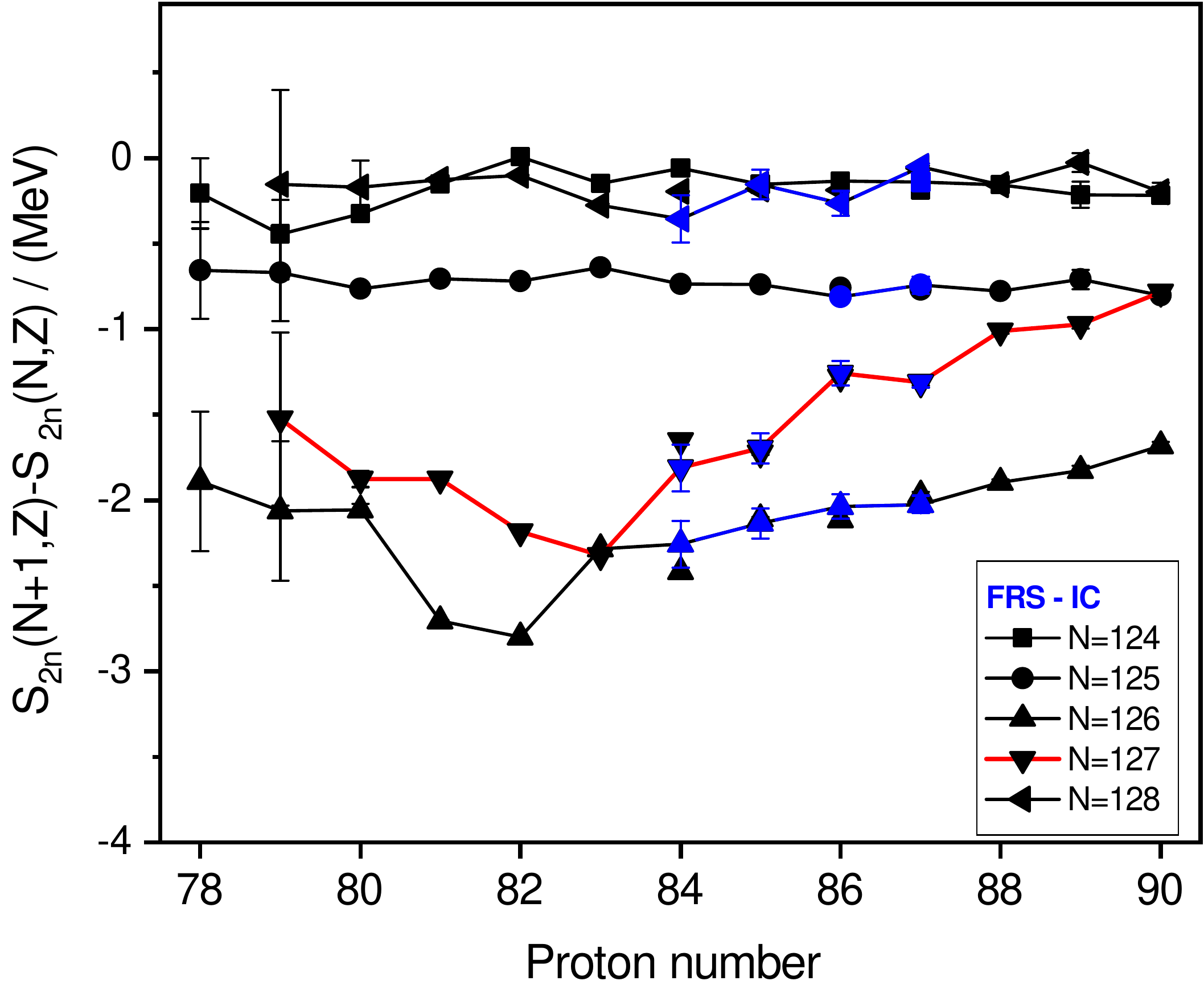}
\caption[Delta 1N S2N]{Slope of the two neutron separation energy $\Delta_\mathrm{1n} S_\mathrm{2n}$ versus proton number. The black symbols show to the data corresponding to the AME16. The red symbols the measurements from this work.}
\label{fig:slope-exp}
\end{figure}

 For this goal, we study the difference given by 
\begin{equation} \label{eq:DeltaS2N_1N}
\Delta_\mathrm{1n} S_\mathrm{2n}\left(N,Z\right) = S_\mathrm{2n}\left(N+1,Z\right) - S_\mathrm{2n}\left(N,Z\right) \; ,
\end{equation}
thus mapping the proton shell closure at lead. Figure \ref{fig:slope-exp} presents 
this correlation using the experimental data of this work and the AME2016 data base \cite{Wang2017}. The symbols denoted by a blue color indicate  the mass excess data of the present experiments.

The observation is that except for the chain of N=126 and N=127 the correlation is very weak, manifested in the horizontal slopes with small staggering. The behavior of N=127 and N=126 is significantly different. The curves exhibit a positive slope beyond the Z=82 shell closure, undergo a minimum
near the proton shell closure. The peculiar experimental observation has been manifested by our direct mass measurements.

In the next step of this study,  it is interesting to investigate the predictions of different theoretical models in the same mass range. We have selected successful mass models based on quite different theoretical
approaches. The predictive power of different models have been recently compared with experimental data in reference \cite{Sobiczewski2014}. 

\begin{figure}[htp]
\centering
\includegraphics[width=0.9\linewidth]{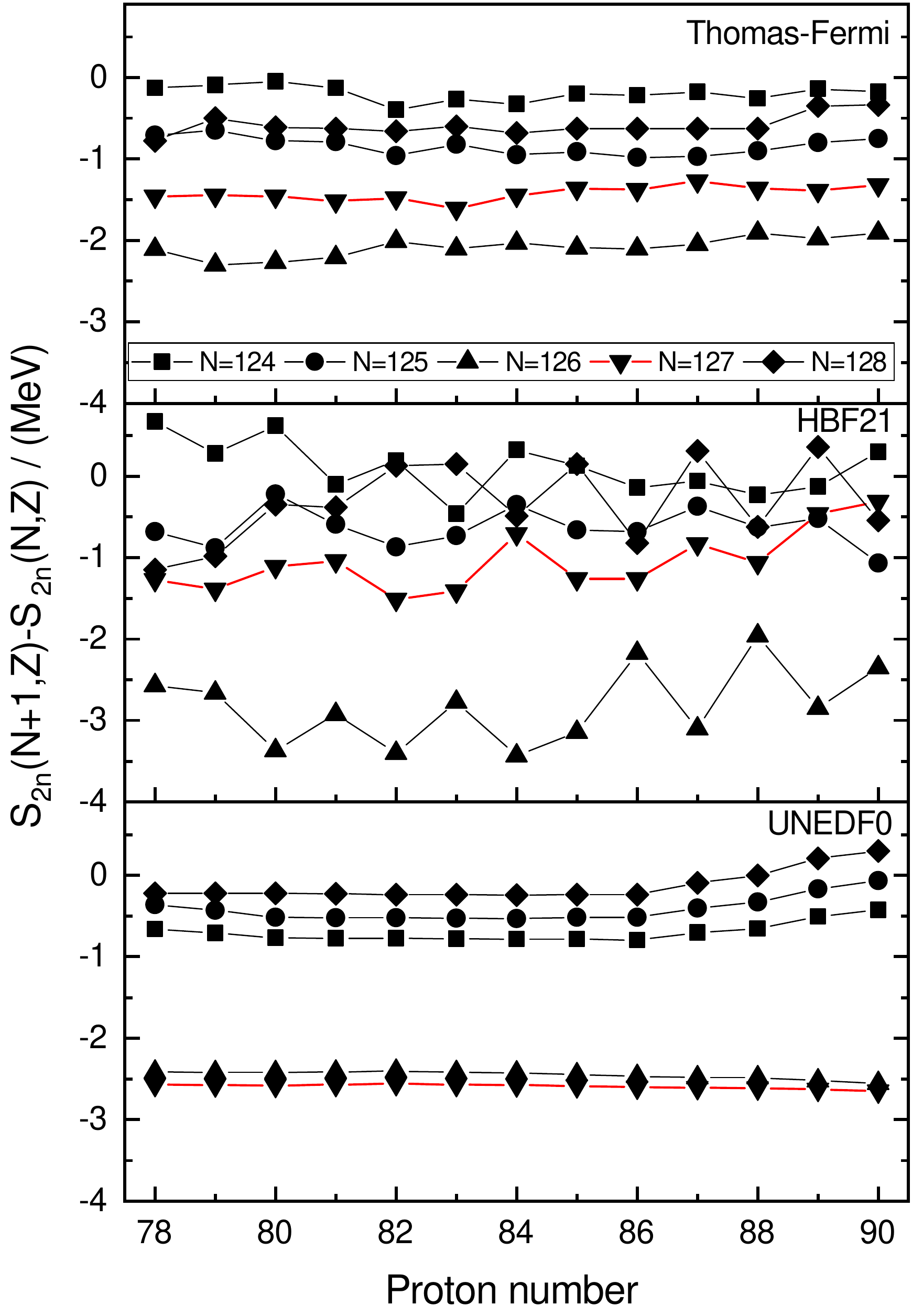}
\caption[Delta 1N S2N theo]{Slope of the two neutron separation energy $\Delta_\mathrm{1n} S_\mathrm{2n}$ versus proton number in different mass models, Thomas-Fermi \cite{MYERS1996}, HBF21 \cite{Goriely2013} and UNDEF0 \cite{Kortelainen2010}.}
\label{fig:slope-theo}
\end{figure}

In figure \ref{fig:slope-theo}, the  theoretical slopes for two neutron separation energies with neutron numbers 124-128 are shown as a function of the atomic number. In general, the theoretical data strongly deviate from the experimental observation, especially for the N=126 and N=127 chains.
In the top panel of figure \ref{fig:slope-theo} the predictions of Myers and Swiatecki \cite{MYERS1996}, based on the Thomas-Fermi (TF) model, are presented. This model
had one of the best predictive power over the full range of isotopes \cite{Sobiczewski2014}. The TF model indicates for all neutron chains horizontal slopes with small staggering features. Except for the N=126 and N=127 chains this reflects the experimental observation. However, for N=126 and N=127 the TF model deviates strongly and has deviations of several hundred keV. 
 The two other theoretical comparisons show two pure microscopic approaches. In the middle panel a Hartree-Fock-Bogolubov model, denoted as 'HFB-21' \cite{Goriely2013} and in the bottom panel an energy density functional 'UNEDF0' \cite{Kortelainen2010} were employed. 
The strong deviation of experimental and theoretical data especially for N=126 and N=127 is manifested in all panels of figure \ref{fig:slope-theo}. 

 In the HFB-21 approach, one can see a staggering with amplitude near 1 MeV, which is clearly not observed in the experimental data. This suggests that the pairing force is not properly adjusted in this model. In the UNEDF0 approach, the corresponding curves are smooth and horizontal. However, both
 theories cannot reproduce the experimental observation. The UNEDF0 model
 has even a reversed order for the N=126 and N=127 shells. 
The observed experimental peculiarity is certainly a stringent test for 
the accuracy and limitation of the present theories. Therefore, it represents 
a strong motivation for theoretical improvements even in the so-called well-known domain of the doubly magic nucleus $^{208}$Pb and for the domain of shell effects of superheavy elements.

\section{Summary and Outlook}

Direct mass measurements of fission and projectile fragments produced with $^{238}$U and $^{124}$Xe primary beams have been performed with the MR-TOF-MS of the FRS Ion Catcher in four experiments. The nuclides were produced, separated in-flight, and energy-bunched with the FRS and finally thermalized in gas-filled cryogenic stopping cell (CSC). An MR-TOF-MS specialized for the accurate mass measurements for rare isotopes, with a few counts only, has been used in the present experiments. A data-analysis procedure was developed to determine the mass values and their uncertainties. The analysis is well suited for overlapping peaks, which solely can be distinguished from a single peak by a change in the peak shape. With this data-analysis procedure, the effective mass resolving power for overlapping peaks is increased by a factor of up to three compared to standard data analysis. This procedure has a direct impact for the resolution of low-lying isomers.

The masses of 31 unstable nuclides with half-lives down to 18~ms were measured. Mass resolving powers beyond 400,000 were achieved. This is the highest mass resolving power reached in mass measurements of short-lived nuclides with an MR-TOF-MS up to now. The masses of six isomeric states with excitation energies down to 280 keV were determined. Nuclides of 15 different elements were measured with count rates as low as 11 events per nuclide. It was further possible to extract mass values for isotopes with ion detection rates of as low as 12 events per hour. The weighted mean of the relative deviations from literature for all the measured masses is $(4.5 \pm 5.3)\cdot 10^{-8}$. The minimum relative uncertainty obtained with the MR-TOF-MS is $6 \cdot 10^{-8}$. 

The first direct mass measurements of seven isotopes close to the double magic nucleus $^{208}$Pb allowed to study the evolution of the two-neutron separation energies. A strong element-dependency is seen for the first neutron above the closed proton shell Z=82. The experimental results deviate strongly from different  theoretical predictions, especially for N=126 and N=127. Therefore, it is a new challenge for the theoretical models even in the so-called well-known domain of the doubly magic nucleus $^{208}$Pb.

These results demonstrate the competitiveness of a high-resolution MR-TOF-MS for measuring masses of short-lived nuclides with an accuracy high enough to yield significant information for nuclear physics and astrophysics. It opens the door for measurements of unknown masses with an accuracy that was up to now only possible with Penning traps.

Future efforts will focus on the following improvements: isobaric continuous calibration will be provided by a newly designed laser ablation carbon cluster ion source included in the upgraded RFQ beamline between the CSC and the \mbox{MR-TOF-MS} \cite{Hornung2018b}; the improved operation mode and electronics for the MRS will make this uncertainty negligible \cite{Bergmann2018b}, and the resolving power will be further increased by means of an improved ion-optical tuning, longer cycle times, and further improved stability of the voltages supplied to the analyzer. With these improvements a relative uncertainty in the range of $2 \cdot 10^{-8}$ is within reach.

While in previous experiments with the FRS Ion Catcher the focus was placed on the commissioning and characterization of the CSC, experiments in the coming years will be dedicated physics experiments.
Several experimental runs will be performed during FAIR Phase-0 at GSI/FAIR. These experiments include mass measurements at $N$=126 below $^{208}$Pb \cite{Pietri2018} and for $N=Z$ nuclides below $^{100}$Sn \cite{Plass2018}. Moreover the system will be used to identify reactions and decay products and thereby measure beta-delayed neutron emission probabilities \cite{Mardor2018} and multi-nucleon transfer reaction product cross-sections \cite{Dickel2018}.

\section*{Acknowledgments}
In memory, we gratefully acknowledge the long-standing, fruitful collaboration with our dear colleague and friend A. Sobiczewski. 

We would like to thank K.-H. Behr, T. Blatz, A. Br{\"u}nle, C. Karagiannis, A. Kratz, C. Lotze, C. Schl{\"o}r, B. Szczepanczky, J. Siebring, T. Wasem and R. Wei{\ss} for excellent technical support. L. Schl{\"u}ter, V. Munoz, G. Kripko-Koncz and M. Macko we would like to thank for helping in debugging the data analysis code. This work was supported by the German Federal Ministry for Education and Research (BMBF) under under contracts no.\ 05P12RGFN8 and 05P15RGFN1, by Justus-Liebig-Universit{\"a}t Gie{\ss}en and GSI under the JLU-GSI strategic Helmholtzpartnership agreement, by HGS-HIRe, and by the Hessian Ministry for Science and Art (HMWK) through the LOEWE Center HICforFAIR.





\bibliographystyle{apsrev4-1}
\bibliography{References.bib}
 






\end{document}